\documentclass[11pt]{article}
\usepackage{amssymb}

\usepackage{amsmath}

\newcounter{resultnum}[section]

\newcounter{conclusionnum}[section]

\newcounter{conditionnum}[section]

\newcounter{conjecturenum}[section]

\newcounter{examplenum}[section]

\newcounter{exercisenum}[section]

\newcounter{lemmanum}[section]

\newcounter{notationnum}[section]

\newcounter{theoremnum}[section]

\newcounter{definitionnum}[section]

\newcounter{corollarynum}[section]

\newcounter{remarknum}[section]

\newcounter{propositionnum}[section]

\newcounter{acknowledgementnum}[section]

\newcounter{algorithmnum}[section]

\newcounter{axiomnum}[section]

\newcounter{casenum}[section]

\newcounter{claimnum}[section]

\newcounter{summarynum}[section]

\newcounter{problemnum}[section]

\begin{document}

\title{Einstein Gravity in Almost K\"{a}hler Variables\\ and Stability of
Gravity with Nonholonomic Distributions and Nonsymmetric Metrics}
\date{November 8, 2008}
\author{ Sergiu I. Vacaru\thanks{%
Sergiu.Vacaru@gmail.com ; http://www.scribd.com/people/view/1455460-sergiu } \\
%EndAName
\textsl{\small The Fields Institute for Research in Mathematical Science} \\
\textsl{\small 222 College Street, 2d Floor, Toronto \ M5T 3J1, Canada} \\
{\small and}\\
\textsl{\small Faculty of Mathematics, University "Al. I. Cuza" Ia\c si }\\
\textsl{\small 700506, Ia\c si, Romania}}
\maketitle

\begin{abstract}
We argue that the Einstein gravity theory can be reformulated in almost 
K\"{a}hler (nonsymmetric) variables with effective symplectic form and compatible linear connection uniquely defined by a (pseudo) Riemannian metric. A class of nonsymmetric theories of gravitation (NGT) on manifolds enabled with nonholonomic distributions is analyzed. There are considered some conditions when the fundamental geometric and physical objects are
determined/ modified by nonholonomic deformations in general relativity or by contributions from Ricci flow theory and/or quantum gravity. We prove that in such NGT, for certain classes of nonholonomic constraints, there are modelled effective Lagrangians which do not develop instabilities. It is
also elaborated a linearization formalism for anholonomic NGT models and analyzed the stability of stationary ellipsoidal solutions defining some nonholonomic and/or nonsymmetric deformations of the Schwarzschild metric.
We show how to construct nonholonomic distributions which remove instabilities in NGT. Finally we conclude that instabilities do not consist a general feature of theories of gravity with nonsymmetric metrics but a particular property of certain models and/or classes of unconstrained solutions.

\vskip3pt


\vskip2pt

PACS2008:\ 04.90.+e, 04.50.+h, 02.40.-k

\vskip1pt

\end{abstract}

\tableofcontents

\section{ Introduction}

In this article, we re--address the issue of nonsymmetric gravity theory
(NGT) following three key ideas: 1) the general relativity theory can be
written equivalently in terms of certain nonsymmetric variables; 2)
nonsymmetric contributions to metrics and connections may be generated in
quasi--classical limits of quantum gravity and nonholonomic and/or
noncommutative Ricci flow theory; 3) physically valuable solutions and their
generalizations with nonsymmetric/ noncommutative / nonholonomic variables
can be stabilized by corresponding classes of nonholonomic constraints on
gravitational field and (geometric) evolution equations. This paper belongs
to a series of three our works on gravity and spaces enabled 
with general symmetric nonsymmetric components metrics and related
nonlinear and linear connection structures, see also partner articles \cite%
{avnsm02,avnsm01}. Our goal is to consider some knew applications in gravity physics and define the conditions when such gravitational "nonsymetric" interactions can be modelled on Einstein spaces.

The Einstein gravity can be represented equivalently in almost K\"{a}hler
(canonical almost symplectic) variables \cite{valmk,vqg4}, see a review of
results in applications of the geometric formalism for constructing exact
solutions in gravity \cite{ijgmmp} and modelling locally anisotropic
interactions in standard theories of physics \cite{ijgmmp2}. Following such
an approach, the data for a (pseudo) Riemannian metric $\mathbf{g}=\{\mathbf{%
g}_{\mu \nu }\}$ and related Levi--Civita connection $\nabla \lbrack \mathbf{%
g}]=\{\ _{\shortmid }\mathbf{\Gamma }_{\ \beta \gamma }^{\alpha }[\mathbf{g}%
]\}$ on a spacetime manifold $\mathbf{V}$ \ (we shall write in brief $(%
\mathbf{g},\nabla ))$ can be equivalently re--defined, in a unique form,
into corresponding almost symplectic form $\mathbf{\theta }=\{\mathbf{\theta
}_{\mu \nu }[\mathbf{g}]\}$ and compatible symplectic connection $\ _{n}%
\mathbf{D[\theta ]}=\{\ _{n}\mathbf{\Gamma }_{\ \beta \gamma }^{\alpha }%
\mathbf{[\theta ]}\}$ (we shall write in brief $(\mathbf{\theta },\ _{n}%
\mathbf{D})),$ for which $\ _{n}\mathbf{D\theta =}\ _{n}\mathbf{Dg=0}.$ The
almost symplectic/ K\"{a}hler connection $\ _{n}\mathbf{D}$ is similar to
the Cartan connection in Finsler--Lagrange geometry \cite%
{opr1,opr2,opr3,mats,ma,vfqlf,avqg5}, but we emphasize that in this article
we shall work only with geometric structures defined on nonholonomic
(pseudo) Riemannian manifolds.\footnote{%
A pair $(\mathbf{V},\mathcal{N})$, where $\mathbf{V}$ is a manifold and $%
\mathcal{N}$ is a nonintegrable distribution on $\mathbf{V}$, is called a
nonholonomic manifold; we note that in our works we use left ''up'' and
''low'' symbols as formal labels for certain geometric objects and that the
spacetime signature may be encoded into formal frame (vielbein)
coefficients, some of them being proportional to the imaginary unity $i,$
when $i^{2}=-1.$}

From a formal point of view, the Cartan's almost symplectic connection
contains nontrivial torsion components induced by the anholonomy
coefficients.\footnote{%
in mathematical and physical literature, there are used also some other
equivalent terms like anholonomic, or non--integrable, restrictions/
constraints; we emphasize that in classical and quantum physics the field
and evolution equations play a fundamental role but together with certain
types of constraints and broken symmetries} But such a nonholonomically
induced torsion is not similar to torsions from the Einstein--Cartan and/or
string/gauge gravity theories, where certain additional field equations (to
the Einstein equations) are considered for torsion fields. The almost K\"{a}%
hler variables are canonically defined for any 2+2 splitting (which allows
us to define canonically an almost complex structure $\mathbf{J})$ in
general relativity, and the induced ''symplectic'' torsion is completely
determined by certain off--diagonal metric components under nonholonomic
deformations of geometric structures. In such cases, $\mathbf{\theta (X,Y)}%
\doteqdot \mathbf{g}\left( \mathbf{J}\mathbf{X,Y}\right) ,$ for any vectors $%
\mathbf{X}$ and $\mathbf{Y}$ on $\mathbf{V},$ and we can consider
equivalently two linear connections subjected to a condition of type $\nabla
\lbrack \mathbf{g}]=\ _{n}\mathbf{D}[\mathbf{g}]+\mathbf{Z}[\mathbf{g}],$
with the distorsion tensor $\mathbf{Z}[\mathbf{g}]$ completely defined by
the original metric field $\mathbf{g.}$

In classical general relativity, it is convenient to work with the variables
$(\mathbf{g},\nabla )$ and, for instance, their tetradic or spinor
representations. For different approaches to quantum gravity, there are
considered $3+1$ spacetime decompositions (for instance, in the so--called
Arnowit--Deser--Misner, ADM, formalism, Ashtekar variables and loop quantum
gravity) or nonholonomic $2+2$ splittings, see a discussion of approaches
and references in \cite{vloopdq}. Even the almost symplectic variables $(%
\mathbf{\theta },\ _{n}\mathbf{D})$ result in a more sophisticate form of
gravitational field equations (similar situations exist for the the ADM
and/or Ashtekar--Barbero representations of gravity), they allow us to apply
directly the deformation quantization formalism and quantize general
relativity following Fedosov's methods \cite{valmk,vqg4}. This is a rigorous
mathematically quantization procedure which provides an alternative approach
to quantum gravity (comparing to various loop, spin--networks methods,
canonical quantization etc methods) even the problem of renormalization of
gravity, if it exists also in a non--perturbative fashion, has not yet been
approached in the Fedosov's theory.

For the almost K\"{a}hler representation of general relativity, the
gravitational symplectic form is anti--symmetric, $\mathbf{\theta }_{\mu \nu
}=-\mathbf{\theta }_{\nu \mu },$ and play the role of ''anti--symmetric''
metric. We can also obtain additional ''nonsymmetric'' metric contributions
from the ''de--quantization'' procedure in deformation quantization of
gravity, or (in a more straightforward form) from the theory of nonholonomic
and/or noncommutatie Ricci flows, see Refs. \cite{vnhrf,avnsm01,vnoncomrf}.
Such geometric quantum constructions and evolution models put in a new
fashion the problem of gravity with nonsymmetric variables. There is already
a long time history, beginning with A. Einstein \cite{nseinst1,nseinst} and
L. P. Eisenhart \cite{nseisenh1,nseisenh2}, when the so--called nonsymmetric
gravity theories have been elaborated in different modifications by J.
Moffat and co--authors \cite{moff1,moffrev,moff1a,nsgtjmp,moff2,moff3,mofft}%
, see also a recent contribution in Ref. \cite{castro}.

A series of works by T. Janssen and T. Prokopec \cite{jp1,jp2,jp3} is
devoted to the so--called ''problem of instabilities'' in NGT. The authors
agreed that one can be elaborated a model of NGT with nonzero mass term for
the nonsymmetric part of metric (treated as an absolutely symmetric torsion
induced by an effective $B$--field like in string gravity, but in four
dimensions). That solved the problems formally created by absence of gauge
invariance found by Damour, Deser and McCarthy \cite{ddmcc}, see explicit
constructions and detailed discussions in \cite{moff1a,cl1}. It was also
emphasized that, as a matter of principle, the Clayton's effect \cite{cl2}
(when, for a general relativity background, a small $B$--field for the
nonsymmetric part quickly grows) may be stabilized by solutions with
evolving backgrounds \cite{isn}\ and/or introducing an extra Lagrange
multiplier when the unstable modes dynamically vanish \cite{moff5}.

Nevertheless, the general conclusion following from works \cite{jp1,jp2,jp3}
is that instabilities in NGT should not be seen as a relic of the linearized
theory because certain nonlinearized NGT models with nontrivial Einstein
background (for instance, on Schwarzschild spacetime) are positively
unstable. Such solutions can not be stabilized by the former methods with
dynamical solutions and, as a consequence, certain new models of NGT and
methods of stabilizations should be developed.

It should be emphasized that the Janssen--Prokopek stability problem does
not have a generic character for all models of gravity with nonsymmetric
variables. As we emphasized above, the Einstein gravity can be represented
equivalently in canonical almost symplectic variables and such a formal
theory with nonsymmetric metric (nonholonomically transformed into
components of a symplectic form) is stable under deformations of the
Schwarzschild metric. But in such a representation, we have also certain
nontrivial nonholonomic structures. So, it is important to study the problem
of stability of physical valuable solutions in general relativity under
nonholonomic deformations, which may keep the constructions in the framework
of the Einstein theory (with certain classes of imposed non--integrable
constraints), or may generalize the gravity theory to models with nontrivial
contributions from Ricci flow evolution (for instance, under variation of
gravitational constants) and/or from a noncommutative/quantum gravity theory.

The goal of this paper is to prove that stable configurations can be derived
for various models of nonsymmetric gravity theories (NGT) \cite%
{moff1,moffrev,moff1a,nsgtjmp,moff2,moff3,mofft}. We shall use a geometric
techniques elaborated in Refs. \cite{ijgmmp2,avnsm01,avnsm02,vrfg,vsgg} and
show how nonholonomic frame constraints can be imposed in order to generate
stable solutions in NGT. For vanishing nonsymmetric components of metrics
such configurations can be reduced to nonholonomic\footnote{%
equivalently, there are used the terms anholonomic and/or nonitegrable} ones
in general relativity (GR) theory and generalizations. We shall provide
explicit examples of stationary solutions with ellipsoidal symmetry which
can be constructed in NGT and GR theories; such metrics are stable and
transform into the Schwarzschild one for zero eccentricities.

In brief, the Janssen--Prokopec method proving that a full, nonlinearized,
NGT may suffer from instabilities can be summarized in this form: One shows
that there is only one stable linearized Lagrangian (see in Ref. \cite{jp1}
the formula (A26) which can be obtained from their formula (86); similar
formulas, (\ref{eflag}) and (\ref{gfspel}), are provided below in Section
3). Then, following certain explicit computations for different backgrounds
in general relativity, one argues that for the Schwarzschild background the
mentioned variant of stable Lagrangian cannot be obtained by linearizing NGT
(because in such cases, the coefficient $\gamma $ in the mentioned formulas,
can not be zero for the static spherical symmetric background in GR).

Generalizing the constructions from \cite{jp1} in order to include certain
types of nonholonomic distributions on (non) symmetric spacetime manifolds,
we shall prove that stable Lagrangians can be generated by a superposition
of nonholonomic transforms and linearization in general models of NGT with
compatible (nonsymmetric) metrics and nonlinear and linear connection
structures.

We argue that fixing from the very beginning an ansatz with spherical
symmetry background (for instance, the Schwarzschild solution in gravity),
one eliminates from consideration a large class of physically important
symmetric and nonsymmetric nonlinear gravitational interactions. The
resulting instability of such constrained to a given background solutions
reflects the proprieties of some very special classes of solutions but not
any intrinsic, fundamental, general characteristics of NGT. For instance, we
shall construct explicit ''ellipsoidal'' stationary solutions in NGT to
which a static Schwarzschild metric is deformed by very small nonsymmetric
metric components and nonholonomic distributions and which seem to be stable
for geometric distorsions in Einstein gravity \cite{vbe1,vbe2}.\footnote{%
in this work, we can consider that the nonsymmetric components of a general
metric induce such geometric and effective matter field distorsions} Such
metrics were constructed for different models of metric--affine, generalized
Finsler on nonholonomic manifolds and noncommutative gravity \cite%
{vncg,ijgmmp,vsgg}) and can be included in NGT both by nonsymmetric metric
components and/or as a nonholonomic symmetric background, see examples from
Ref. \cite{avnsm01}.

The paper is organized as follows: In Section 2, we outline some basic
results from the geometry of nonholonomic manifolds and NGT models on such
spaces. The equivalent formulation of the Einstein gravity in canonical
almost symplectic variables is provided. Section 3 is devoted to a method of
nonholonomic deformations and linearization to backgrounds with symmetric
metrics and nonholonomic distributions. We show how certain classes of
nonsymmetric metric configurations can be stabilized by corresponding
nonholonomic constraints. We present an explicit example in Section 4, when
stable stationary solutions with nontrivial nonsymmetric components of
metric and nonholonomic distributions are constructed as certain
deformations of the Schwarzschild metric to an ellipsoidal nonholonomic
background on which a constrained dynamics on nonsymmetric metric fields is
modelled. Finally, in Section 5 we present conclusions and discuss the
results. In Appendix, we provide some important formulas on torsion and
curvature of linear connections adapted to a prescribed nonlinear connection
structure.

\section{Einstein Gravity in Almost K\"{a}hler Variables}

In general relativity (GR), we consider a real four dimensional (pseudo)
Riemanian spacetime manifold $V$ of signature $(-,+,+,+)$ and necessary
smooth class. For a conventional $2+2$ splitting, the local coordinates $%
u=(x,y)$ on a open region $U\subset V$ are labelled in the form $u^{\alpha
}=(x^{i},y^{a}),$ where indices of type $i,j,k,...=1,2$ and $a,b,c...=3,4,$
for tensor like objects, will be considered with respect to a general
(non--coordinate) local basis $e_{\alpha }=(e_{i},e_{a}).$ One says that $%
x^{i}$ and $y^{a}$ are respectively the conventional horizontal/ holonomic
(h) and vertical / nonholonomic (v) coordinates (both types of such
coordinates can be time-- or space--like ones). \ Primed indices of type $%
i^{\prime },a^{\prime },...$ will be used for labelling coordinates with
respect to a different local basis $e_{\alpha ^{\prime }}=(e_{i^{\prime
}},e_{a^{\prime }})$ or $e_{\alpha ^{\prime }}=(e_{0}^{\prime },e_{I^{\prime
}}),$ for instance, for an orthonormalized basis. For the local tangent
Minkowski space, we chose $e_{0^{\prime }}=i\partial /\partial u^{0^{\prime
}},$ where $i$ is the imaginary unity, $i^{2}=-1,$ and write $e_{\alpha
^{\prime }}=(i\partial /\partial u^{0^{\prime }},\partial /\partial
u^{1^{\prime }},\partial /\partial u^{2^{\prime }},\partial /\partial
u^{3^{\prime }}).$ To consider such formal Euclidean coordinates is useful
for some purposes of analogous modelling of gravity theories as effective
Lagrange mechanics geometries, but this does not mean that we introduce any
complexification of classical spacetimes. In this section, we outline the
constructions for classical gravity from \cite{valmk,vqg4,vfqlf}.

\subsection{N--anholonomic (pseudo) Riemannian manifolds}

The coefficients of a general (pseudo) Riemannian metric on a spacetime $V$
\ are parametrized in the form: \
\begin{eqnarray}
\mathbf{g} &=&g_{i^{\prime }j^{\prime }}(u)e^{i^{\prime }}\otimes
e^{j^{\prime }}+h_{a^{\prime }b^{\prime }}(u)e^{a^{\prime }}\otimes
e^{b^{\prime }},  \label{gpsm} \\
e^{a^{\prime }} &=&\mathbf{e}^{a^{\prime }}-N_{i^{\prime }}^{a^{\prime
}}(u)e^{i^{\prime }},  \notag
\end{eqnarray}%
where the required form of vierbein coefficients $e_{\ \alpha }^{\alpha
^{\prime }}$ of the dual basis
\begin{equation}
e^{\alpha ^{\prime }}=(e^{i^{\prime }},e^{a^{\prime }})=e_{\ \alpha
}^{\alpha ^{\prime }}(u)du^{\alpha },  \label{dft}
\end{equation}%
defining a formal $2+2$ splitting, will be stated below.

On spacetime $V,$ we consider any generating function $L(u)=L(x^{i},y^{a})$
(we may call it a formal Lagrangian if an effective continuous mechanical
model of GR is to be elaborated, see Refs. \cite{ijgmmp2,vrfg,vsgg}) with
nondegenerate Hessian
\begin{equation}
\ ^{L}h_{ab}=\frac{1}{2}\frac{\partial ^{2}L}{\partial y^{a}\partial y^{b}},
\label{elm}
\end{equation}%
when $\det |\ ^{L}h_{ab}|\neq 0.$ This function is useful for constructing
in explicit form a nonholonomic $2+2$ splitting for which a canonical almost
symplectic model of GR will be defined. We use $L$ as an abstract label and
emphasize that the geometric constructions are general ones, not depending
on the type of function $L(u)$ which states only a formal class of systems
of reference and coordinates. \ Working with such local fibrations, it is a
more simple procedure to define almost symplectic variables in GR. We
introduce
\begin{equation}
\ ^{L}N_{i}^{a}=\frac{\partial G^{a}}{\partial y^{2+i}},  \label{clnc}
\end{equation}%
for
\begin{equation}
G^{a}=\frac{1}{4}\ ^{L}h^{a\ 2+i}\left( \frac{\partial ^{2}L}{\partial
y^{2+i}\partial x^{k}}y^{2+k}-\frac{\partial L}{\partial x^{i}}\right) ,
\label{clssp}
\end{equation}%
where $\ ^{L}h^{ab}$ is inverse to $\ ^{L}h_{ab}$ and respective
contractions of $h$-- and $v$--indices, $\ i,j,...$ and $a,b...,$ are
performed following the rule: we can write, for instance, an up $v$--index $%
a $ as $a=2+i$ and contract it with a low index $i=1,2.$ Briefly, we shall
write $y^{i}$ instead of $y^{2+i},$ or $y^{a}.$ The values (\ref{elm}), (\ref%
{clnc}) and (\ref{clssp}) allow us to define
\begin{eqnarray}
^{L}\mathbf{g} &=&\ ^{L}g_{ij}dx^{i}\otimes dx^{j}+\ ^{L}h_{ab}\ ^{L}\mathbf{%
e}^{a}\otimes \ ^{L}\mathbf{e}^{b},  \label{lfsm} \\
\ ^{L}\mathbf{e}^{a} &=&dy^{a}+\ ^{L}N_{i}^{a}dx^{i},\ ^{L}g_{ij}=\
^{L}h_{2+i\ 2+j}.  \notag
\end{eqnarray}

A metric $\mathbf{g}$ (\ref{gpsm}) with coefficients $g_{\alpha ^{\prime
}\beta ^{\prime }}=[g_{i^{\prime }j^{\prime }},h_{a^{\prime }b^{\prime }}]$
computed with respect to a dual basis $e^{\alpha ^{\prime }}=(e^{i^{\prime
}},e^{a^{\prime }})$ can be related to the metric $\ ^{L}\mathbf{g}_{\alpha
\beta }=\ [\ ^{L}g_{ij},\ ^{L}h_{ab}]$ (\ref{lfsm}) with coefficients
defined with respect to a N--adapted dual basis $\ ^{L}e^{\alpha }=(dx^{i},\
^{L}\mathbf{e}^{a})$ if there are satisfied the conditions%
\begin{equation}
\mathbf{g}_{\alpha ^{\prime }\beta ^{\prime }}e_{\ \alpha }^{\alpha ^{\prime
}}e_{\ \beta }^{\beta ^{\prime }}=~^{L}\mathbf{g}_{\alpha \beta }.
\label{algeq}
\end{equation}%
Considering any given values $\mathbf{g}_{\alpha ^{\prime }\beta ^{\prime }}$
and $~^{L}\mathbf{g}_{\alpha \beta },$ we have to solve a system of
quadratic algebraic equations with unknown variables $e_{\ \alpha }^{\alpha
^{\prime }}.$ How to define locally such coordinates, we discuss in Ref. %
\cite{ijgmmp2,vloopdq}. For instance, in GR, there are 6 independent values $%
\mathbf{g}_{\alpha ^{\prime }\beta ^{\prime }}$ and up till ten coefficients
$~^{L}\mathbf{g}_{\alpha \beta }$ which allows us always to define a set of
vierbein coefficients $e_{\ \alpha }^{\alpha ^{\prime }}.$ Usually, a subset
of such coefficients can be taken be zero, for given values $[g_{i^{\prime
}j^{\prime }},h_{a^{\prime }b^{\prime }},N_{i^{\prime }}^{a^{\prime }}]$ and
$\ [\ ^{L}g_{ij},\ ^{L}h_{ab},\ \ ^{L}N_{i}^{a}],$ when
\begin{equation}
N_{i^{\prime }}^{a^{\prime }}=e_{i^{\prime }}^{\ i}e_{\ a}^{a^{\prime }}\ \
^{L}N_{i}^{a}  \label{defcncl}
\end{equation}%
for $e_{i^{\prime }}^{\ i}$ being inverse to $e_{\ i}^{i^{\prime }}.$

For simplicity, in this work, we suppose that there is always a finite
covering of $\mathbf{V}^{2+2}$ (in brief, denoted $\mathbf{V})$ by a family
of open regions $\ ^{I}U,$ labelled by an index $I,$ on which there are
considered certain nontrivial effective Lagrangians $\ ^{I}L$ with real
solutions $\ ^{I}e_{\ \alpha }^{\alpha ^{\prime }}$ defining vielbein
transforms to systems of so--called Lagrange variables. Finally, we solve
the algebraic equations (\ref{algeq}) for any prescribed values $%
g_{i^{\prime }j^{\prime }}$ (we also have to change the partition $\ ^{I}U$
and generating function $\ ^{I}L$ till we are able to construct real
solutions) and find $\ ^{I}e_{\ i}^{i^{\prime }}$ which, in its turn, allows
us to compute $N_{i^{\prime }}^{a^{\prime }}$(\ref{defcncl}) and all
coefficients of the metric $\mathbf{g}$ (\ref{gpsm}) and vierbein transform (%
\ref{dft}). We shall omit for simplicity the left labe $L$ if that will not
result in a confusion for some special constructions.

A nonlinear connection (N--connection) structure $\mathbf{N}$ for $\mathbf{V}
$ is defined by a nonholonomic distribution (a Whitney sum)%
\begin{equation}
T\mathbf{V}=h\mathbf{V}\oplus v\mathbf{V}  \label{whitney}
\end{equation}%
into conventional horizontal (h) and vertical (v) subspaces. In local form,
a N--connection is given by its coefficients $N_{i}^{a}(u),$ when%
\begin{equation}
\mathbf{N}=N_{i}^{a}(u)dx^{i}\otimes \frac{\partial }{\partial y^{a}}.
\label{coeffnc}
\end{equation}

A N--connection introduces on $\mathbf{V}^{n+n}$ a frame (vielbein)
structure
\begin{equation}
\mathbf{e}_{\nu }=\left( \mathbf{e}_{i}=\frac{\partial }{\partial x^{i}}%
-N_{i}^{a}(u)\frac{\partial }{\partial y^{a}},e_{a}=\frac{\partial }{%
\partial y^{a}}\right) ,  \label{dder}
\end{equation}%
and a dual frame (coframe) structure
\begin{equation}
\mathbf{e}^{\mu }=\left( e^{i}=dx^{i},\mathbf{e}%
^{a}=dy^{a}+N_{i}^{a}(u)dx^{i}\right) .  \label{ddif}
\end{equation}%
The vielbeins (\ref{ddif}) satisfy the nonholonomy relations
\begin{equation}
\lbrack \mathbf{e}_{\alpha },\mathbf{e}_{\beta }]=\mathbf{e}_{\alpha }%
\mathbf{e}_{\beta }-\mathbf{e}_{\beta }\mathbf{e}_{\alpha }=w_{\alpha \beta
}^{\gamma }\mathbf{e}_{\gamma }  \label{anhrel}
\end{equation}%
with (antisymmetric) nontrivial anholonomy coefficients $w_{ia}^{b}=\partial
_{a}N_{i}^{b}$ and $w_{ji}^{a}=\Omega _{ij}^{a},$ where
\begin{equation}
\Omega _{ij}^{a}=\mathbf{e}_{j}\left( N_{i}^{a}\right) -\mathbf{e}_{i}\left(
N_{j}^{a}\right)  \label{ncurv}
\end{equation}%
are the coefficients of N--connection curvature (defined as the Neijenhuis
tensor on $\mathbf{V}^{n+n}).$ The particular holonomic/ integrable case is
selected by the integrability conditions $w_{\alpha \beta }^{\gamma }=0.$%
\footnote{%
we use boldface symbols for spaces (and geometric objects on such spaces)
enabled with N--connection structure}

A N--anholonomic manifold is a (nonholonomic) manifold enabled with
N--connection structure (\ref{whitney}). The geometric properties of a
N--anholonomic manifold are distinguished by some N--adapted bases (\ref%
{dder}) and (\ref{ddif}). A geometric object is N--adapted (equivalently,
distinguished), i.e. a d--object, if it can be defined by components adapted
to the splitting (\ref{whitney}) (one uses terms d--vector, d--form,
d--tensor). For instance, a d--vector $\mathbf{X}=X^{\alpha }\mathbf{e}%
_{\alpha }=X^{i}\mathbf{e}_{i}+X^{a}e_{a}$ and a one d--form $\widetilde{%
\mathbf{X}}$ (dual to $\mathbf{X}$) is $\widetilde{\mathbf{X}}=X_{\alpha }%
\mathbf{e}^{\alpha }=X_{i}e^{i}+X_{a}e^{a}.$\footnote{%
We can redefine equivalently the geometric constructions for arbitrary frame
and coordinate systems; the N--adapted constructions allow us to preserve
the h-- and v--splitting.}

\subsection{Canonical almost symplectic structures in GR}

Let $\mathbf{e}_{\alpha ^{\prime }}=(\mathbf{e}_{i},e_{b^{\prime }})$ and $%
\mathbf{e}^{\alpha ^{\prime }}=(e^{i},\ \mathbf{e}^{b^{\prime }})$ be
defined respectively by (\ref{dder}) and (\ref{ddif}) for the canonical
N--connection $\ ^{L}\mathbf{N}$ (\ref{clnc}) stated by a metric structure $%
\mathbf{g}=$ $~^{L}\mathbf{g}$ (\ref{lfsm}) on $\mathbf{V}.$ We introduce a
linear operator $\mathbf{J}$ acting on vectors on $\mathbf{V}$ following
formulas
\begin{equation*}
\mathbf{J}(\mathbf{e}_{i})=-e_{2+i}\mbox{\ and \ }\mathbf{J}(e_{2+i})=%
\mathbf{e}_{i},
\end{equation*}%
where and $\mathbf{J\circ J=-I}$ for $\mathbf{I}$ being the unity matrix,
and construct a tensor field on $\mathbf{V},$%
\begin{eqnarray}
\mathbf{J} &=&\mathbf{J}_{\ \beta }^{\alpha }\ e_{\alpha }\otimes e^{\beta }=%
\mathbf{J}_{\ \underline{\beta }}^{\underline{\alpha }}\ \frac{\partial }{%
\partial u^{\underline{\alpha }}}\otimes du^{\underline{\beta }}
\label{acstr} \\
&=&\mathbf{J}_{\ \beta ^{\prime }}^{\alpha ^{\prime }}\ \mathbf{e}_{\alpha
^{\prime }}\otimes \mathbf{e}^{\beta ^{\prime }}=\mathbf{-}e_{2+i}\otimes
e^{i}+\mathbf{e}_{i}\otimes \ \mathbf{e}^{2+i}  \notag \\
&=&-\frac{\partial }{\partial y^{i}}\otimes dx^{i}+\left( \frac{\partial }{%
\partial x^{i}}-\ ^{L}N_{i}^{2+j}\frac{\partial }{\partial y^{j}}\right)
\otimes \left( dy^{i}+\ ^{L}N_{k}^{2+i}dx^{k}\right) ,  \notag
\end{eqnarray}%
defining globally an almost complex structure on\ $\mathbf{V}$ completely
determined by a fixed $L(x,y).$ Using vielbeins $\mathbf{e}_{\ \underline{%
\alpha }}^{\alpha }$ and their duals $\mathbf{e}_{\alpha \ }^{\ \underline{%
\alpha }}$, defined by $e_{\ \alpha }^{\alpha ^{\prime }}$ solving (\ref%
{algeq}), we can compute the coefficients of tensor $\mathbf{J}$ with
respect to any local basis $e_{\alpha }$ and $e^{\alpha }$ on $\mathbf{V},$ $%
\mathbf{J}_{\ \beta }^{\alpha }=\mathbf{e}_{\ \underline{\alpha }}^{\alpha }%
\mathbf{J}_{\ \underline{\beta }}^{\underline{\alpha }}\mathbf{e}_{\beta \
}^{\ \underline{\beta }}.$ In general, we can define an almost complex
structure $\mathbf{J}$ for an arbitrary N--connection $\mathbf{N,}$ stating
a nonholonomic $2+2$ splitting, by using N--adapted bases (\ref{dder}) and (%
\ref{ddif}).

The Neijenhuis tensor field for any almost complex structure $\mathbf{J}$
defined by a N--connection (equivalently, the curvature of N--connecti\-on)
is
\begin{equation}
\ ^{\mathbf{J}}\mathbf{\Omega (X,Y)\doteqdot -[X,Y]+[JX,JY]-J[JX,Y]-J[X,JY],}
\label{neijt}
\end{equation}%
for any d--vectors $\mathbf{X}$ and $\mathbf{Y.}$ With respect to N--adapted
bases (\ref{dder}) and (\ref{ddif}), a subset of the coefficients of the
Neijenhuis tensor defines the N--connection curvature, see details in Ref. %
\cite{ma},
\begin{equation}
\Omega _{ij}^{a}=\frac{\partial N_{i}^{a}}{\partial x^{j}}-\frac{\partial
N_{j}^{a}}{\partial x^{i}}+N_{i}^{b}\frac{\partial N_{j}^{a}}{\partial y^{b}}%
-N_{j}^{b}\frac{\partial N_{i}^{a}}{\partial y^{b}}.  \label{nccurv}
\end{equation}%
A N--anholonomic manifold $\mathbf{V}$ is integrable if $\Omega _{ij}^{a}=0.$
We get a complex structure if and only if both the h-- and v--distributions
are integrable, i.e. if and only if $\Omega _{ij}^{a}=0$ and $\frac{\partial
N_{j}^{a}}{\partial y^{i}}-\frac{\partial N_{i}^{a}}{\partial y^{j}}=0.$

One calls an almost symplectic structure on a manifold $V$ a nondegenerate
2--form
\begin{equation*}
\theta =\frac{1}{2}\theta _{\alpha \beta }(u)e^{\alpha }\wedge e^{\beta }.
\end{equation*}%
For any $\theta $ on $V,$ there is a unique N--connection $\mathbf{N}%
=\{N_{i}^{a}\}$ (\ref{whitney}) satisfying the conditions:%
\begin{equation}
\theta =(h\mathbf{X},v\mathbf{Y})=0\mbox{ and }\theta =h\theta +v\theta ,
\label{aux02}
\end{equation}%
for any $\mathbf{X}=h\mathbf{X}+v\mathbf{X,}$ $\mathbf{Y}=h\mathbf{Y}+v%
\mathbf{Y},$ where $h\theta (\mathbf{X,Y})\doteqdot \theta (h\mathbf{X,}h%
\mathbf{Y})$ and $v\theta (\mathbf{X,Y})\doteqdot \theta (v\mathbf{X,}v%
\mathbf{Y}).$

For $\mathbf{X=e}_{\alpha }=(\mathbf{e}_{i},e_{a})$ and $\mathbf{Y=e}_{\beta
}=(\mathbf{e}_{l},e_{b}),$ where $\mathbf{e}_{\alpha }$ is a N--adapted
basis\ of type (\ref{dder}), we write the first equation in (\ref{aux02}) in
the form%
\begin{equation*}
\theta =\theta (\mathbf{e}_{i},e_{a})=\theta (\frac{\partial }{\partial x^{i}%
},\frac{\partial }{\partial y^{a}})-N_{i}^{b}\theta (\frac{\partial }{%
\partial y^{b}},\frac{\partial }{\partial y^{a}})=0.
\end{equation*}%
We can solve this system of equations in a unique form and define $N_{i}^{b}$
if $rank|\theta (\frac{\partial }{\partial y^{b}},\frac{\partial }{\partial
y^{a}})|=2.$ Denoting locally
\begin{equation}
\mathbf{\theta }=\frac{1}{2}\theta _{ij}(u)e^{i}\wedge e^{j}+\frac{1}{2}%
\theta _{ab}(u)\mathbf{e}^{a}\wedge \mathbf{e}^{b},  \label{aux03}
\end{equation}%
where the first term is for $h\theta $ and the second term is $v\theta ,$ we
get the second formula in (\ref{aux02}).

An almost Hermitian model of a (pseudo) Riemannian spa\-ce $\mathbf{V}$
equipped with a N--connection structure $\mathbf{N}$ is defined by a triple $%
\mathbf{H}^{2+2}=(\mathbf{V},\theta ,\mathbf{J}),$ where $\mathbf{\theta
(X,Y)}\doteqdot \mathbf{g}\left( \mathbf{JX,Y}\right) $ for any $\mathbf{g}$
(\ref{gpsm}). A space $\mathbf{H}^{2+2}$ is almost K\"{a}hler, denoted $%
\mathbf{K}^{2+2},$ if and only if $d\mathbf{\theta }=0.$

For $\mathbf{g}=\ $ $^{L}\mathbf{g}$ (\ref{lfsm}) and structures $\ ^{L}%
\mathbf{N}$ (\ref{clnc}) and $\mathbf{J}$ canonically defined by $L,$ we
define $\ ^{L}\mathbf{\theta (X,Y)}\doteqdot \ ^{L}\mathbf{g}\left( \mathbf{%
JX,Y}\right) $ for any d--vectors $\mathbf{X}$ and $\mathbf{Y.}$ In local
N--adapted form form, we have
\begin{eqnarray}
\ ^{L}\mathbf{\theta } &=&\frac{1}{2}\ ^{L}\theta _{\alpha \beta
}(u)e^{\alpha }\wedge e^{\beta }=\frac{1}{2}\ ^{L}\theta _{\underline{\alpha
}\underline{\beta }}(u)du^{\underline{\alpha }}\wedge du^{\underline{\beta }}
\label{asymstr} \\
&=&\ ^{L}g_{ij}(x,y)e^{2+i}\wedge dx^{j}=\ ^{L}g_{ij}(x,y)(dy^{2+i}+\
^{L}N_{k}^{2+i}dx^{k})\wedge dx^{j}.  \notag
\end{eqnarray}%
Let us consider the form $\ ^{L}\omega =\frac{1}{2}\frac{\partial L}{%
\partial y^{i}}dx^{i}.$ A straightforward computation shows that $\ ^{L}%
\mathbf{\theta }=d\ ^{L}\omega ,$ which means that $d\ ^{L}\mathbf{\theta }%
=dd\ ^{L}\omega =0,$ i.e. the canonical effective Lagrange structures $%
\mathbf{g}=\ ^{L}\mathbf{g},\ \ ^{L}\mathbf{N}$ and $\mathbf{J}$ induce an
almost K\"{a}hler geometry. We can express the 2--form (\ref{asymstr}) as
\begin{eqnarray}
\mathbf{\theta } &=&\ ^{L}\mathbf{\theta }=\frac{1}{2}\ ^{L}\theta
_{ij}(u)e^{i}\wedge e^{j}+\frac{1}{2}\ ^{L}\theta _{ab}(u)\mathbf{e}%
^{a}\wedge \mathbf{e}^{b}  \label{canalmsf} \\
&=&g_{ij}(x,y)\left[ dy^{i}+N_{k}^{i}(x,y)dx^{k}\right] \wedge dx^{j},
\notag
\end{eqnarray}%
see (\ref{aux03}), where the coefficients $\ ^{L}\theta _{ab}=\ ^{L}\theta
_{2+i\ 2+j}$ are equal respectively to the coefficients $\ ^{L}\theta _{ij}.$
It should be noted that for a general 2--form $\theta $ constructed for any
metric $\mathbf{g}$ and almost complex $\mathbf{J}$\textbf{\ }structures on $%
V$ one holds $d\theta \neq 0.$ But for any $2+2$ splitting induced by an
effective Lagrange generating function, we have $d\ ^{L}\mathbf{\theta }=0.$
We have also $d\ \mathbf{\theta }=0$ for any set of 2--form coefficients $%
\mathbf{\theta }_{\alpha ^{\prime }\beta ^{\prime }}e_{\ \alpha }^{\alpha
^{\prime }}e_{\ \beta }^{\beta ^{\prime }}=~^{L}\mathbf{\theta }_{\alpha
^{\prime }\beta ^{\prime }}$ (such a 2--form $\mathbf{\theta }$ will be
called to be a canonical one), constructed by using formulas (\ref{algeq}).

We conclude that having chosen a generating function $L(x,y)$ on a (pseudo)
Riemannian spacetime $\mathbf{V},$ we can model this spacetime equivalently
as an almost K\"{a}hler manifold

\subsection{Equivalent metric compatible linear connections}

A distinguished connection (in brief, d--connection) on a spacetime $\mathbf{%
V}$,
\begin{equation*}
\mathbf{D}=(hD;vD)=\{\mathbf{\Gamma }_{\beta \gamma }^{\alpha
}=(L_{jk}^{i},\ ^{v}L_{bk}^{a};C_{jc}^{i},\ ^{v}C_{bc}^{a})\},
\end{equation*}%
is a linear connection which preserves under parallel transports the
distribution (\ref{whitney}). In explicit form, the coefficients\ $\mathbf{%
\Gamma }_{\beta \gamma }^{\alpha }$ are computed with respect to a
N--adapted basis (\ref{dder}) and (\ref{ddif}). A d--connection $\mathbf{D}$%
\ is metric compatible with a d--metric $\mathbf{g}$ if $\mathbf{D}_{\mathbf{%
X}}\mathbf{g}=0$ for any d--vector field $\mathbf{X.}$

If an almost symplectic structure $\theta $ is considered on a
N--anholonomic manifold, an almost symplectic d--connection $\ _{\theta }%
\mathbf{D}$ on $\mathbf{V}$ is defined by the conditions that it is
N--adapted, i.e. it is a d--connection, and $\ _{\theta }\mathbf{D}_{\mathbf{%
X}}\theta =0,$ for any d--vector $\mathbf{X.}$ From the set of metric and/or
almost symplectic compatible d--connecti\-ons on a (pseudo) Riemannian
manifold $\mathbf{V},$ we can select those which are completely defined by a
metric $\mathbf{g}=\ $ $^{L}\mathbf{g}$ (\ref{lfsm}) and an effective
Lagrange structure $L(x,y):$

There is a unique normal d--connection
\begin{eqnarray}
\ \widehat{\mathbf{D}} &=&\left\{ h\widehat{D}=(\widehat{D}_{k},^{v}\widehat{%
D}_{k}=\widehat{D}_{k});v\widehat{D}=(\widehat{D}_{c},\ ^{v}\widehat{D}_{c}=%
\widehat{D}_{c})\right\}  \label{ndc} \\
&=&\{\widehat{\mathbf{\Gamma }}_{\beta \gamma }^{\alpha }=(\widehat{L}%
_{jk}^{i},\ ^{v}\widehat{L}_{2+j\ 2+k}^{2+i}=\widehat{L}_{jk}^{i};\ \widehat{%
C}_{jc}^{i}=\ ^{v}\widehat{C}_{2+j\ c}^{2+i},\ ^{v}\widehat{C}_{bc}^{a}=%
\widehat{C}_{bc}^{a})\},  \notag
\end{eqnarray}%
which is metric compatible,
\begin{equation*}
\widehat{D}_{k}\ ^{L}g_{ij}=0\mbox{ and }\widehat{D}_{c}\ ^{L}g_{ij}=0,
\end{equation*}%
and completely defined by a couple of h-- and v--components $\ \widehat{%
\mathbf{D}}_{\alpha }=(\widehat{D}_{k},\widehat{D}_{c}),$ with N--adapted
coefficients $\widehat{\mathbf{\Gamma }}_{\beta \gamma }^{\alpha }=(\widehat{%
L}_{jk}^{i},\ ^{v}\widehat{C}_{bc}^{a}),$ where
\begin{eqnarray}
\widehat{L}_{jk}^{i} &=&\frac{1}{2}\ ^{L}g^{ih}\left( \mathbf{e}_{k}\
^{L}g_{jh}+\mathbf{e}_{j}\ ^{L}g_{hk}-\mathbf{e}_{h}\ ^{L}g_{jk}\right) ,
\label{cdcc} \\
\widehat{C}_{jk}^{i} &=&\frac{1}{2}\ ^{L}g^{ih}\left( \frac{\partial \
^{L}g_{jh}}{\partial y^{k}}+\frac{\partial \ ^{L}g_{hk}}{\partial y^{j}}-%
\frac{\partial \ ^{L}g_{jk}}{\partial y^{h}}\right) .  \notag
\end{eqnarray}%
In general, we can ''foget'' about label $L$ and work with arbitrary $%
\mathbf{g}_{\alpha ^{\prime }\beta ^{\prime }}$ and $\widehat{\mathbf{\Gamma
}}_{\beta ^{\prime }\gamma ^{\prime }}^{\alpha ^{\prime }}$ with the
coefficients recomputed by frame transforms (\ref{dft}).

Introducing the normal d--connection 1--form%
\begin{equation*}
\widehat{\mathbf{\Gamma }}_{j}^{i}=\widehat{L}_{jk}^{i}e^{k}+\widehat{C}%
_{jk}^{i}\mathbf{e}^{k},
\end{equation*}%
we prove that the Cartan structure equations are satisfied,%
\begin{equation}
de^{k}-e^{j}\wedge \widehat{\mathbf{\Gamma }}_{j}^{k}=-\widehat{\mathcal{T}}%
^{i},\ d\mathbf{e}^{k}-\mathbf{e}^{j}\wedge \widehat{\mathbf{\Gamma }}%
_{j}^{k}=-\ ^{v}\widehat{\mathcal{T}}^{i},  \label{cart1}
\end{equation}%
and
\begin{equation}
d\widehat{\mathbf{\Gamma }}_{j}^{i}-\widehat{\mathbf{\Gamma }}_{j}^{h}\wedge
\widehat{\mathbf{\Gamma }}_{h}^{i}=-\widehat{\mathcal{R}}_{\ j}^{i}.
\label{cart2}
\end{equation}%
The h-- and v--components of the torsion 2--form
\begin{equation*}
\widehat{\mathcal{T}}^{\alpha }=\left( \widehat{\mathcal{T}}^{i},\ ^{v}%
\widehat{\mathcal{T}}^{i}\right) =\widehat{\mathbf{T}}_{\ \tau \beta
}^{\alpha }\ \mathbf{e}^{\tau }\wedge \mathbf{e}^{\beta }
\end{equation*}%
from (\ref{cart1}) is computed with components
\begin{equation}
\widehat{\mathcal{T}}^{i}=\widehat{C}_{jk}^{i}e^{j}\wedge \mathbf{e}^{k},\
^{v}\widehat{\mathcal{T}}^{i}=\frac{1}{2}\ ^{L}\Omega _{kj}^{i}e^{k}\wedge
e^{j}+(\frac{\partial \ \ ^{L}N_{k}^{i}}{\partial y^{j}}-\widehat{L}_{\
kj}^{i})e^{k}\wedge \mathbf{e}^{j},  \label{tform}
\end{equation}%
where $\ ^{L}\Omega _{kj}^{i}$ are coefficients of the curvature of the
canonical N--connection $\check{N}_{k}^{i}$ defined by formulas similar to (%
\ref{nccurv}). The formulas (\ref{tform}) parametrize the h-- and
v--components of torsion $\widehat{\mathbf{T}}_{\beta \gamma }^{\alpha }$ in
the form
\begin{equation}
\widehat{T}_{jk}^{i}=0,\widehat{T}_{jc}^{i}=\widehat{C}_{\ jc}^{i},\widehat{T%
}_{ij}^{a}=\ ^{L}\Omega _{ij}^{a},\widehat{T}_{ib}^{a}=e_{b}\left( \
^{L}N_{i}^{a}\right) -\widehat{L}_{\ bi}^{a},\widehat{T}_{bc}^{a}=0.
\label{cdtors}
\end{equation}%
It should be noted that $\widehat{\mathbf{T}}$ vanishes on h- and
v--subspaces, i.e. $\widehat{T}_{jk}^{i}=0$ and $\widehat{T}_{bc}^{a}=0,$
but certain nontrivial h--v--components induced by the nonholonomic
structure are defined canonically by $\mathbf{g}=\ ^{L}\mathbf{g}$ (\ref%
{lfsm}) and $L.$ For convenience, in Appendix \ref{append1}, we outline some
important component formulas for the canonical d--connection which on spaces
of even dimensions transform into those for the normal connection.

We compute also the curvature 2--form from (\ref{cart2}),%
\begin{eqnarray}
\widehat{\mathcal{R}}_{\ \gamma }^{\tau } &=&\widehat{\mathbf{R}}_{\ \gamma
\alpha \beta }^{\tau }\ \mathbf{e}^{\alpha }\wedge \ \mathbf{e}^{\beta }
\label{cform} \\
&=&\frac{1}{2}\widehat{R}_{\ jkh}^{i}e^{k}\wedge e^{h}+\widehat{P}_{\
jka}^{i}e^{k}\wedge \mathbf{e}^{a}+\frac{1}{2}\ \widehat{S}_{\ jcd}^{i}%
\mathbf{e}^{c}\wedge \mathbf{e}^{d},  \notag
\end{eqnarray}%
where the nontrivial N--adapted coefficients of curvature $\widehat{\mathbf{R%
}}_{\ \beta \gamma \tau }^{\alpha }$ of $\widehat{\mathbf{D}}$ are
\begin{eqnarray}
\widehat{R}_{\ hjk}^{i} &=&\mathbf{e}_{k}\widehat{L}_{\ hj}^{i}-\mathbf{e}%
_{j}\widehat{L}_{\ hk}^{i}+\widehat{L}_{\ hj}^{m}\widehat{L}_{\ mk}^{i}-%
\widehat{L}_{\ hk}^{m}\widehat{L}_{\ mj}^{i}-\widehat{C}_{\ ha}^{i}\
^{L}\Omega _{\ kj}^{a}  \label{cdcurv} \\
\widehat{P}_{\ jka}^{i} &=&e_{a}\widehat{L}_{\ jk}^{i}-\widehat{\mathbf{D}}%
_{k}\widehat{C}_{\ ja}^{i},  \notag \\
\widehat{S}_{\ bcd}^{a} &=&e_{d}\widehat{C}_{\ bc}^{a}-e_{c}\widehat{C}_{\
bd}^{a}+\widehat{C}_{\ bc}^{e}\widehat{C}_{\ ed}^{a}-\widehat{C}_{\ bd}^{e}%
\widehat{C}_{\ ec}^{a}.  \notag
\end{eqnarray}%
Contracting the first and forth indices $\widehat{\mathbf{\mathbf{R}}}%
\mathbf{_{\ \beta \gamma }=}\widehat{\mathbf{\mathbf{R}}}\mathbf{_{\ \beta
\gamma \alpha }^{\alpha }}$, we get the N--adapted coefficients for the
Ricci tensor%
\begin{equation}
\widehat{\mathbf{\mathbf{R}}}\mathbf{_{\beta \gamma }=}\left( \widehat{R}%
_{ij},\widehat{R}_{ia},\widehat{R}_{ai},\widehat{R}_{ab}\right) .
\label{dricci}
\end{equation}%
The scalar curvature $\ ^{L}R=\widehat{R}$ of $\widehat{\mathbf{D}}$ is
\begin{equation}
\ ^{L}R=\ ^{L}\mathbf{g}^{\beta \gamma }\widehat{\mathbf{\mathbf{R}}}\mathbf{%
_{\beta \gamma }=\ \mathbf{g}^{\beta ^{\prime }\gamma ^{\prime }}\widehat{%
\mathbf{\mathbf{R}}}\mathbf{_{\beta ^{\prime }\gamma ^{\prime }}}.}
\label{scalc}
\end{equation}

The normal d--connection $\widehat{\mathbf{D}}$ (\ref{ndc}) defines a
canonical almost symplectic d--connection, $\widehat{\mathbf{D}}\equiv \
_{\theta }\widehat{\mathbf{D}},$ which is N--adapted to the effective
Lagrange and, related, almost symplectic structures, i.e. it preserves under
parallelism the splitting (\ref{whitney}), $_{\theta }\widehat{\mathbf{D}}_{%
\mathbf{X}}\ ^{L}\mathbf{\theta =}_{\theta }\widehat{\mathbf{D}}_{\mathbf{X}%
}\ \mathbf{\theta =}0$ and its torsion is constrained to satisfy the
conditions $\widehat{T}_{jk}^{i}=\widehat{T}_{bc}^{a}=0.$

In the canonical approach to the general relativity theory, one works with
the Levi Civita connection $\bigtriangledown =\{\ _{\shortmid }\Gamma
_{\beta \gamma }^{\alpha }\}$ which is uniquely derived following the
conditions $~\ _{\shortmid }\mathcal{T}=0$ and $\bigtriangledown \mathbf{g}%
=0.$ This is a linear connection but not a d--connection because $%
\bigtriangledown $ does not preserve (\ref{whitney}) under parallelism. Both
linear connections $\bigtriangledown $ and $\widehat{\mathbf{D}}\equiv \
_{\theta }\widehat{\mathbf{D}}$ are uniquely defined in metric compatible
forms by the same metric structure $\mathbf{g}$ (\ref{gpsm}). The second one
contains nontrivial d--torsion components $\widehat{\mathbf{T}}_{\beta
\gamma }^{\alpha }$ (\ref{cdtors}), induced effectively by an equivalent
Lagrange metric $\mathbf{g}=\ ^{L}\mathbf{g}$ (\ref{lfsm}) and adapted both
to the N--connection $\ ^{L}\mathbf{N,}$ see (\ref{clnc}) and (\ref{whitney}%
), and almost symplectic $\ ^{L}\mathbf{\theta }$ (\ref{asymstr}) structures
$L.$

Any geometric construction for the normal d--connection $\widehat{\mathbf{D}}%
(\mathbf{\theta })$ can be re--defined by the Levi Civita connection, and
inversely, using the formula
\begin{equation}
\ _{\shortmid }\Gamma _{\ \alpha \beta }^{\gamma }(\mathbf{\theta })=%
\widehat{\mathbf{\Gamma }}_{\ \alpha \beta }^{\gamma }(\mathbf{\theta })+\
_{\shortmid }Z_{\ \alpha \beta }^{\gamma }(\mathbf{\theta }),  \label{cdeft}
\end{equation}%
where the both connections $\ _{\shortmid }\Gamma _{\ \alpha \beta }^{\gamma
}(\mathbf{\theta })$ and $\widehat{\mathbf{\Gamma }}_{\ \alpha \beta
}^{\gamma }(\mathbf{\theta })$ and the distorsion tensor $\ _{\shortmid
}Z_{\ \alpha \beta }^{\gamma }(\mathbf{g})$ with N--adapted coefficients
(for the normal d--connection $\ _{\shortmid }Z_{\ \alpha \beta }^{\gamma }(%
\mathbf{g})$ is proportional to $\widehat{\mathbf{T}}_{\beta \gamma
}^{\alpha }(\mathbf{g})$ (\ref{cdtors})), see formulas (\ref{deft}). In this
work, we emphasize if it is necessary the functional dependence of certain
geometric objects on a d--metric $(\mathbf{g}),$ or its canonical almost
symplectic equivalent $(\mathbf{\theta })$\ for tensors and connections
completely defined by the metric structure.\footnote{%
see Appendix on similar deformation properties of fundamental geometric
objects}

If we work with nonholonomic constraints on the dynamics/ geometry of
gravity fields in deformation quantization, it is more convenient to use a
N--adapted and/or almost symplectic approach. For other purposes, it is
preferred to use only the Levi--Civita connection. Introducing the
distorsion relation (\ref{cdeft}) into respective formulas (\ref{cdtors}), (%
\ref{cdcurv}) and (\ref{dricci}) written for $\widehat{\mathbf{\Gamma }}_{\
\alpha \beta }^{\gamma },$ we get deformations
\begin{eqnarray}
\ _{\shortmid }T_{\ \beta \gamma }^{\alpha }(\mathbf{g}) &=&\widehat{\mathbf{%
T}}_{\ \beta \gamma }^{\alpha }(\mathbf{g})+\ _{\shortmid }Z_{\ \alpha \beta
}^{\gamma }(\mathbf{g})=0,  \label{aux01} \\
\ _{\shortmid }R_{\ \beta \gamma \delta }^{\alpha }(\mathbf{g}) &=&\widehat{%
\mathbf{R}}_{\ \beta \gamma \delta }^{\alpha }+\ _{\shortmid }\widehat{%
\mathbf{Z}}_{\ \beta \gamma \delta }^{\alpha }(\mathbf{g}),\ _{\shortmid
}R_{\ \beta \gamma }(\mathbf{g})=\widehat{\mathbf{R}}_{\ \beta \gamma }+\
_{\shortmid }\widehat{\mathbf{Z}}_{\ \beta \gamma }(\mathbf{g}),  \notag
\end{eqnarray}%
see Refs. \cite{vrfg,vsgg} for explicit formulas for distorisons of the
torsion, curvature, Ricci tensors, i.e. $\ _{\shortmid }^{T}Z_{\ \alpha
\beta }^{\gamma }(\mathbf{g}),$ $\ _{\shortmid }\widehat{\mathbf{Z}}_{\
\beta \gamma \delta }^{\alpha }(\mathbf{g})$ and $\ _{\shortmid }\widehat{%
\mathbf{Z}}_{\ \beta \gamma }(\mathbf{g}),$ which are completely defined by
a metric structure $\mathbf{g}=\ ^{L}\mathbf{g}$ with a nonholonomic 2+2
splitting induced by a prescribed regular $L.$ Such formulas can be
re--defined equivalently for $\ _{\shortmid }^{T}Z_{\ \alpha \beta }^{\gamma
}(\mathbf{\theta }),$ $\ _{\shortmid }\widehat{\mathbf{Z}}_{\ \beta \gamma
\delta }^{\alpha }(\mathbf{\theta })$ and $\ _{\shortmid }\widehat{\mathbf{Z}%
}_{\ \beta \gamma }(\mathbf{\theta }),$ written only in terms of the
canonical almost symplectic from $\mathbf{\theta }$ (\ref{canalmsf}).

\subsection{An almost symplectic formulation of GR}

\label{ssakgr}Having chosen a canonical almost symplectic d--connection, we
compute the Ricci d--tensor $\widehat{\mathbf{R}}_{\ \beta \gamma }$ (\ref%
{dricci}) and the scalar curvature $\ ^{L}R$ $\ $(\ref{scalc})). Then, we
can postulate in a straightforward form the filed equations
\begin{equation}
\widehat{\mathbf{R}}_{\ \beta }^{\underline{\alpha }}-\frac{1}{2}(\
^{L}R+\lambda )\mathbf{e}_{\ \beta }^{\underline{\alpha }}=8\pi G\mathbf{T}%
_{\ \beta }^{\underline{\alpha }},  \label{deinsteq}
\end{equation}%
where $\widehat{\mathbf{R}}_{\ \ \beta }^{\underline{\alpha }}=\mathbf{e}_{\
\gamma }^{\underline{\alpha }}$ $\widehat{\mathbf{R}}_{\ \ \beta }^{\ \gamma
},$ $\mathbf{T}_{\ \beta }^{\underline{\alpha }}$ is the effective
energy--momentum tensor, $\lambda $ is the cosmological constant, $G$ is the
Newton constant in the units when the light velocity $c=1,$ and the
coefficients $\mathbf{e}_{\ \beta }^{\underline{\alpha }}$ of vierbein
decomposition $\mathbf{e}_{\ \beta }=\mathbf{e}_{\ \beta }^{\underline{%
\alpha }}\partial /\partial u^{\underline{\alpha }}$ are defined by the
N--coefficients of the N--elongated operator of partial derivation, see (\ref%
{dder}). But the equations (\ref{deinsteq}) for the canonical $\widehat{%
\mathbf{\Gamma }}_{\ \alpha \beta }^{\gamma }(\mathbf{\theta })$ are not
equivalent to the Einstein equations in GR writen for the Levi--Civita
connection $_{\shortmid }\Gamma _{\ \alpha \beta }^{\gamma }(\mathbf{\theta }%
)$ if the tensor $\mathbf{T}_{\ \beta }^{\underline{\alpha }}$ does not
include contributions of $\ _{\shortmid }Z_{\ \alpha \beta }^{\gamma }(%
\mathbf{\theta })$ in a necessary form.

Introducing the absolute antisymmetric tensor $\epsilon _{\alpha \beta
\gamma \delta }$ and the effective source 3--form
\begin{equation*}
\mathcal{T}_{\ \beta }=\mathbf{T}_{\ \beta }^{\underline{\alpha }}\ \epsilon
_{\underline{\alpha }\underline{\beta }\underline{\gamma }\underline{\delta }%
}du^{\underline{\beta }}\wedge du^{\underline{\gamma }}\wedge du^{\underline{%
\delta }}
\end{equation*}%
and expressing the curvature tensor $\widehat{\mathcal{R}}_{\ \gamma }^{\tau
}=\widehat{\mathbf{R}}_{\ \gamma \alpha \beta }^{\tau }\ \mathbf{e}^{\alpha
}\wedge \ \mathbf{e}^{\beta }$ of $\ \widehat{\mathbf{\Gamma }}_{\ \beta
\gamma }^{\alpha }=\ _{\shortmid }\Gamma _{\ \beta \gamma }^{\alpha }-\ \
_{\shortmid }\widehat{\mathbf{Z}}_{\ \beta \gamma }^{\alpha }$ as $\widehat{%
\mathcal{R}}_{\ \gamma }^{\tau }=\ _{\shortmid }\mathcal{R}_{\ \gamma
}^{\tau }-\ _{\shortmid }\widehat{\mathcal{Z}}_{\ \gamma }^{\tau },$ where $%
\ _{\shortmid }\mathcal{R}_{\ \gamma }^{\tau }$ $=\ _{\shortmid }R_{\ \gamma
\alpha \beta }^{\tau }\ \mathbf{e}^{\alpha }\wedge \ \mathbf{e}^{\beta }$ is
the curvature 2--form of the Levi--Civita connection $\nabla $ and the
distorsion of curvature 2--form $\widehat{\mathcal{Z}}_{\ \gamma }^{\tau }$
is defined by $\ \widehat{\mathbf{Z}}_{\ \beta \gamma }^{\alpha },$ see (\ref%
{cdeft}) and (\ref{aux01}), we derive the equations (\ref{deinsteq})
(varying the action on components of $\mathbf{e}_{\ \beta },$ see details in
Ref. \cite{vloopdq}). The gravitational field equations are represented as
3--form equations,%
\begin{equation}
\epsilon _{\alpha \beta \gamma \tau }\left( \mathbf{e}^{\alpha }\wedge
\widehat{\mathcal{R}}^{\beta \gamma }+\lambda \mathbf{e}^{\alpha }\wedge \
\mathbf{e}^{\beta }\wedge \ \mathbf{e}^{\gamma }\right) =8\pi G\mathcal{T}%
_{\ \tau },  \label{einsteq}
\end{equation}%
when%
\begin{eqnarray*}
\mathcal{T}_{\ \tau } &=&\ ^{m}\mathcal{T}_{\ \tau }+\ ^{Z}\widehat{\mathcal{%
T}}_{\ \tau }, \\
\ ^{m}\mathcal{T}_{\ \tau } &=&\ ^{m}\mathbf{T}_{\ \tau }^{\underline{\alpha
}}\epsilon _{\underline{\alpha }\underline{\beta }\underline{\gamma }%
\underline{\delta }}du^{\underline{\beta }}\wedge du^{\underline{\gamma }%
}\wedge du^{\underline{\delta }}, \\
\ ^{Z}\mathcal{T}_{\ \tau } &=&\left( 8\pi G\right) ^{-1}\widehat{\mathcal{Z}%
}_{\ \tau }^{\underline{\alpha }}\epsilon _{\underline{\alpha }\underline{%
\beta }\underline{\gamma }\underline{\delta }}du^{\underline{\beta }}\wedge
du^{\underline{\gamma }}\wedge du^{\underline{\delta }},
\end{eqnarray*}%
where $\ ^{m}\mathbf{T}_{\ \tau }^{\underline{\alpha }}$ is the matter
tensor field. The above mentioned equations are equivalent to the usual
Einstein equations for the Levi--Civita connection $\nabla ,$%
\begin{equation*}
\ _{\shortmid }\mathbf{R}_{\ \beta }^{\underline{\alpha }}-\frac{1}{2}(\
_{\shortmid }R+\lambda )\mathbf{e}_{\ \beta }^{\underline{\alpha }}=8\pi G\
^{m}\mathbf{T}_{\ \beta }^{\underline{\alpha }}.
\end{equation*}

The vacuum Einstein equations with cosmological constant, written in terms
of the canonical N--adapted vierbeins and normal d--connection, are%
\begin{equation}
\epsilon _{\alpha \beta \gamma \tau }\left( \mathbf{e}^{\alpha }\wedge
\widehat{\mathcal{R}}^{\beta \gamma }+\lambda \mathbf{e}^{\alpha }\wedge
\mathbf{e}^{\beta }\wedge \ \mathbf{e}^{\gamma }\right) =8\pi G\ ^{Z}%
\widehat{\mathcal{T}}_{\ \tau },  \label{veinst1}
\end{equation}%
with effective source $\ ^{Z}\widehat{\mathcal{T}}_{\ \tau }$ induced by
nonholonomic splitting by the metric tensor and its off--diagonal components
transformed into the N--connection coefficients or, in terms of the
Levi--Civita connection%
\begin{equation*}
\epsilon _{\alpha \beta \gamma \tau }\left( \mathbf{e}^{\alpha }\wedge \
_{\shortmid }\mathcal{R}^{\beta \gamma }+\lambda \mathbf{e}^{\alpha }\wedge
\mathbf{e}^{\beta }\wedge \ \mathbf{e}^{\gamma }\right) =0.
\end{equation*}%
Such formulas\ expressed in terms of canonical almost symplectic form $%
\mathbf{\theta }$ (\ref{canalmsf}) and normal d--connection $\widehat{%
\mathbf{D}}\equiv \ _{\theta }\widehat{\mathbf{D}}$ (\ref{ndc}) are
necessary for encoding the vacuum field equations into cohomological
structure of quantum almost K\"{a}hler models of the Einstein gravity, see %
\cite{valmk,vqg4,vfqlf,avqg5,vloopdq}.

If former geometric constructions in GR were related to frame and coordinate
form invariant transforms, various purposes in geometric modelling of
physical interactions and quantization request application of more general
classes of transforms. For such generalizations, the linear connection
structure is deformed (in a unique/canonical form following well defined
geometric and physical principles) and there are considered nonholonomic
spacetime distributions. All geometric and physical information for any data
1) $[\mathbf{g,}\ _{\shortmid }\Gamma _{\ \alpha \beta }^{\gamma }(\mathbf{g}%
)]$ are transformed equivalently for canonical constructions with 2) $[%
\mathbf{g}=\ ^{L}\mathbf{g,}\ \mathbf{N},$ $\ \widehat{\mathbf{\Gamma }}_{\
\alpha \beta }^{\gamma }(\ \mathbf{g})],$ which allows us to provide an
effective Lagrange interpretation of the Einstein gravity, or 3) $[\ \mathbf{%
\theta }=\ ^{L}\mathbf{\theta },\ _{\theta }\widehat{\mathbf{\Gamma }}_{\
\alpha \beta }^{\gamma }=\widehat{\mathbf{\Gamma }}_{\ \alpha \beta
}^{\gamma },\mathbf{J(\ \theta )}],$ for an almost K\"{a}hler model of
general relativity. The canonical almost symplectic form $\mathbf{\theta }$ (%
\ref{canalmsf}) represents the ''original'' metric $\mathbf{g}$ (\ref{gpsm})
equivalently in a ''nonsymmetric'' form. Any deformations of such
structures, in the framework of GR or quantized models and generalizations,
result in more general classes of nonsymmetric metrics.

\section{NGT with Nonholonomic Distributions}

In this section, we follow the geometric conventions and results from Ref. %
\cite{avnsm02}. The aim is to outline some basic definitions, concepts and
formulas from the geometry of nonholonomic manifolds enabled with nonlinear
connection and general nonsymmetric structure and introduce a general
Lagrangian for NGT and corresponding nonholonomic distributions.

\subsection{Preliminaries: geometry of N--anholonomic manifolds}

In this paper, we also consider gravity models on spaces $(\check{g}_{ij},%
\mathbf{V}^{n+n},\mathbf{N})$ when the h--subspace is enabled with a
nonsymmetric tensor field (metric) $\check{g}_{ij}=g_{ij}+a_{ij},$ where the
symmetric part $g_{ij}=g_{ji}$ is nondegenerated and $a_{ij}=-a_{ji}.$ A
d--metric $\check{g}_{ij}(x,y)$ is of index $k$ if there are satisfied the
properties: 1. $\det |g_{ij}|\neq 0$ and 2. $rank|a_{ij}|=n-k=2p,$ for $%
0\leq k\leq n.$ By $g^{ij}$ we note the reciprocal (inverse) to $g_{ij}$
d--tensor field. The matrix $g_{ij}$ is not invertible unless for $k=0.$

We write by $g^{ij}$ the reciprocal (inverse) to $g_{ij}$ d--tensor field.
The matrix $g_{ij}$ is not invertible unless for $k=0.$ For $k>0$ and a
positive definite $g_{ij}(x,y),$ on each domain of local chart there exists $%
k$ d--vector fields $\xi _{i^{\prime }}^{i},$ where $i=1,2,...,n$ and $%
i^{\prime }=1,...,k$ with the properties%
\begin{equation*}
a_{ij}\xi _{j^{\prime }}^{j}=0\mbox{ and }g_{ij}\xi _{i^{\prime }}^{i}\xi
_{j^{\prime }}^{j}=\delta _{i^{\prime }j^{\prime }}.
\end{equation*}%
If $g_{ij}$ is not positive definite, we shall assume the existence of $k$
linearly independent d--vector fields with such properties.

The metric properties on $\mathbf{V}^{n+n}$ are supposed to be defined by
d--tensor
\begin{eqnarray}
\mathbf{\check{g}} &=&\mathbf{g}+\mathbf{a}=\mathbf{\check{g}}_{\alpha \beta
}\mathbf{e}^{\alpha }\otimes \mathbf{e}^{\beta }=\check{g}_{ij}e^{i}\otimes
e^{j}+\check{g}_{ab}\mathbf{e}^{a}\otimes \mathbf{e}^{b},  \label{hvm} \\
\mathbf{g} &=&\mathbf{g}_{\alpha \beta }\mathbf{e}^{\alpha }\otimes \mathbf{e%
}^{\beta }=g_{ij}e^{i}\otimes e^{j}+g_{ab}\mathbf{e}^{a}\otimes \mathbf{e}%
^{b},  \label{shvm} \\
\mathbf{a} &=&a_{ij}e^{i}\wedge e^{j}+a_{cb}\mathbf{e}^{c}\wedge \mathbf{e}%
^{b},  \notag
\end{eqnarray}%
where the v--components $\check{g}_{ab}$ are defined by the same
coefficients as $\check{g}_{ij}.$ With respect to a coordinate local cobasis
$du^{\alpha }=(dx^{i},dy^{a}),$ we have equivalently
\begin{equation*}
\mathbf{g}=\underline{g}_{\alpha \beta }du^{\alpha }\otimes du^{\beta },
\end{equation*}%
where
\begin{equation}
\underline{g}_{\alpha \beta }=\left[
\begin{array}{cc}
g_{ij}+N_{i}^{a}N_{j}^{b}g_{ab} & N_{j}^{e}g_{ae} \\
N_{i}^{e}g_{be} & g_{ab}%
\end{array}%
\right] .  \label{ansatz}
\end{equation}

A h--v--metric on a N--anholonomic manifold is a second rank d--tensor of
type (\ref{hvm}). We can define the local d--covector fields $\eta
_{i}^{i^{\prime }}=g_{ij}\xi _{i^{\prime }}^{j}$ and the d--tensors of type $%
(1,1),l_{j}^{i}$ and $m_{j}^{i},$ satisfying the conditions%
\begin{eqnarray*}
\mathit{l}_{j}^{i} &=&\xi _{i^{\prime }}^{i}\eta _{j}^{i^{\prime }}%
\mbox{
and }\mathit{m}_{j}^{i}=\delta _{j}^{i}-\xi _{i^{\prime }}^{i}\eta
_{j}^{i^{\prime }},\mbox{ for  }i^{\prime }=1,...,k; \\
\mathit{l}_{j}^{i} &=&0\mbox{ and }\mathit{m}_{j}^{i}=\delta _{j}^{i},%
\mbox{
for  }k=0.
\end{eqnarray*}%
One considers the matrices
\begin{eqnarray}
\widehat{g} &=&(g_{ij}),\widehat{a}=(a_{ij}),\widehat{\xi }=(\xi _{i^{\prime
}}^{i}),\widehat{\mathit{l}}=(\mathit{l}_{j}^{i}),  \label{matr01} \\
\widehat{\eta } &=&(\eta _{i}^{i^{\prime }}),\widehat{\mathit{m}}=(\mathit{m}%
_{j}^{i}),\widehat{\delta }^{\prime }=(\delta _{i^{\prime }j^{\prime }}),%
\widehat{\delta }=(\delta _{j}^{i}).  \notag
\end{eqnarray}

The next step is to extend the matrix $\widehat{a}$ to a nonsingular skew
symmetric one of dimension $(n+k,n+k),$%
\begin{equation*}
\widetilde{a}=\left[
\begin{array}{cc}
\widehat{a} & -\ ^{t}\varphi \\
\varphi & 0%
\end{array}%
\right] .
\end{equation*}%
The inverse matrix $\widetilde{a}^{-1},$ satisfying the condition $%
\widetilde{a}\widetilde{a}^{-1}=\widehat{\delta },$ has the form
\begin{equation}
\widetilde{a}^{-1}=\left[
\begin{array}{cc}
\check{a} & \widehat{\xi } \\
\ ^{t}\widehat{\xi } & 0%
\end{array}%
\right] ,  \label{aux00}
\end{equation}%
where the matrix $\check{a}=\left( \check{a}^{ij}\right) $ does not depend
on the choice of $\widehat{\xi }$ \ and it is uniquely defined by $\widehat{a%
}\check{a}=\ ^{t}\widehat{\mathit{m}}$ and $\widehat{\mathit{l}}\check{a}=0,$
i.e. this matrix is uniquely defined on $\mathbf{V}^{n+n}.$

In general, the concept of linear connection (adapted or not adapted to a
N--connection structure) is independent from the concept of metric
(symmetric or nonsymmetric). A distinguished connection (d--connection) $%
\mathbf{D}$ on $\mathbf{V}$ is a N--adapted linear connection, preserving by
parallelism the vertical and horizontal distribution (\ref{whitney}). In
local form, $\mathbf{D=}\left( \ ^{h}D,\ ^{v}D\right) $ is given by its
coefficients $\mathbf{\Gamma }_{\ \alpha \beta }^{\gamma }=\left(
L_{jk}^{i},L_{bk}^{a},C_{jc}^{i},C_{bc}^{a}\right) ,$ where $\
^{h}D=(L_{jk}^{i},L_{bk}^{a})$ and $\ ^{v}D=(C_{jc}^{i},C_{bc}^{a})$ are
respectively the covariant h-- and v--derivatives. For any d--connection, we
can compute the torsion, curvature and Ricci tensors and scalar curvature,
see Appendix.

A normal d--connection $\ _{n}\mathbf{D}$ is compatible with the almost
complex structure \ $\mathbf{J}$ (\ref{acstr}), i.e. satisfies the condition
$\ $%
\begin{equation}
\ _{n}\mathbf{D}_{\mathbf{X}}\mathbf{J}=0,  \label{acscomp}
\end{equation}%
for any d--vector $\mathbf{X}$ on $\mathbf{V}^{n+n}.$ The operator $_{n}%
\mathbf{D}$ is characterized by a pair of local coefficients $\ _{n}\mathbf{%
\Gamma }_{\ \alpha \beta }^{\gamma }=\left( \ _{n}L_{jk}^{i},\
_{n}C_{bc}^{a}\right) $ defined by conditions
\begin{eqnarray*}
\ _{n}\mathbf{D}_{\mathbf{e}_{k}}(\mathbf{e}_{j}) &=&\ \ _{n}L_{jk}^{i}%
\mathbf{e}_{i},\ \ _{n}\mathbf{D}_{\mathbf{e}_{k}}(e_{a})=\
_{n}L_{ak}^{b}e_{b}: \\
&&\mbox{for }j=a,i=b,\ \ _{n}L_{jk}^{i}=\ \ _{n}L_{ak}^{b}, \\
\ \ _{n}\mathbf{D}_{e_{c}}(\mathbf{e}_{j}) &=&\ \ _{n}C_{jc}^{i}\mathbf{e}%
_{i},\ \ _{n}\mathbf{D}_{e_{c}}(e_{a})=\ \ _{n}C_{ac}^{b}e_{b}: \\
&&\mbox{ for }j=a,i=b,\ _{n}C_{jc}^{i}=\ _{n}C_{ac}^{b}.
\end{eqnarray*}%
Here we emphazie that the normal d--connection $_{n}\mathbf{D}$ is different
from $\widehat{\mathbf{D}}$ (\ref{ndc}) (the first one is defined for a
space with nonsymmetric metrics, but for the second one the metrics must by
symmetric).

A d--connection $\mathbf{D}=\{\mathbf{\Gamma }_{\ \alpha \beta }^{\gamma }\}$
is compatible with a nonsymmetric d--metric $\mathbf{\check{g}}$ if
\begin{equation}
\mathbf{D}_{k}\check{g}_{ij}=0\mbox{ and }\mathbf{D}_{a}\check{g}_{ij}=0.
\label{nscomp}
\end{equation}%
For a d--metric (\ref{hvm}), the equations (\ref{nscomp}) are
\begin{equation}
\mathbf{D}_{k}g_{ij}=0,\mathbf{D}_{a}g_{bc}=0,\mathbf{D}_{k}a_{ij}=0,\mathbf{%
D}_{e}a_{bc}=0.  \label{nscompg}
\end{equation}%
The set of d--connections $\{\mathbf{D}\}$ satisfying the conditions $%
\mathbf{D}_{\mathbf{X}}\mathbf{g=0}$ for a given $\mathbf{g}$ is defined by
formulas%
\begin{eqnarray*}
L_{\ jk}^{i} &=&\widehat{L}_{jk}^{i}+\ ^{-}O_{km}^{ei}\mathbf{X}%
_{ej}^{m\,},\ L_{\ bk}^{a}=\widehat{L}_{bk}^{a}+\ ^{-}O_{bd}^{ca}\mathbf{Y}%
_{ck}^{d\,}, \\
C_{\ jc}^{i} &=&\widehat{C}_{jc}^{i}+\ ^{+}O_{jk}^{mi}\mathbf{X}%
_{mc}^{k\,},\ C_{\ bc}^{a}=\widehat{C}_{bc}^{a}+\ ^{+}O_{bd}^{ea}\mathbf{Y}%
_{ec}^{d\,},
\end{eqnarray*}%
where
\begin{equation}
\ ^{\pm }O_{jk}^{ih}=\frac{1}{2}(\delta _{j}^{i}\delta _{k}^{h}\pm
g_{jk}g^{ih}),\ ^{\pm }O_{bd}^{ca}=\frac{1}{2}(\delta _{b}^{c}\delta
_{d}^{a}\pm g_{bd}g^{ca})  \label{obop}
\end{equation}%
are the so--called the Obata operators; $\mathbf{X}_{ej}^{m\,},\mathbf{X}%
_{mc}^{k\,},\mathbf{Y}_{ck}^{d\,}$ and $\mathbf{Y}_{ec}^{d\,}$ are arbitrary
d--tensor fields and $\widehat{\mathbf{\Gamma }}_{\ \alpha \beta }^{\gamma
}=\left( \widehat{L}_{jk}^{i},\widehat{L}_{bk}^{a},\widehat{C}_{jc}^{i},%
\widehat{C}_{bc}^{a}\right) ,$ with
\begin{eqnarray}
\widehat{L}_{jk}^{i} &=&\frac{1}{2}g^{ir}\left(
e_{k}g_{jr}+e_{j}g_{kr}-e_{r}g_{jk}\right) ,  \label{candcon} \\
\widehat{L}_{bk}^{a} &=&e_{b}(N_{k}^{a})+\frac{1}{2}g^{ac}\left(
e_{k}g_{bc}-g_{dc}\ e_{b}N_{k}^{d}-g_{db}\ e_{c}N_{k}^{d}\right) ,  \notag \\
\widehat{C}_{jc}^{i} &=&\frac{1}{2}g^{ik}e_{c}g_{jk},\ \widehat{C}_{bc}^{a}=%
\frac{1}{2}g^{ad}\left( e_{c}g_{bd}+e_{c}g_{cd}-e_{d}g_{bc}\right)  \notag
\end{eqnarray}%
is the canonical d--connections uniquely defined by the coefficients of
d--metric $\mathbf{g=}[g_{ij},g_{ab}]$ and N--connection $\mathbf{N}%
=\{N_{i}^{a}\}$ in order to satisfy the conditions $\widehat{\mathbf{D}}_{%
\mathbf{X}}\mathbf{g=0}$ and $\widehat{T}_{\ jk}^{i}=0$ and $\widehat{T}_{\
bc}^{a}=0$ but $\widehat{T}_{\ ja}^{i},\widehat{T}_{\ ji}^{a}$ and $\widehat{%
T}_{\ bi}^{a}$ are not zero (on definition of torsion, see formulas (\ref%
{dtors}) in Appendix; we can compute the torsion coefficients $\widehat{%
\mathbf{T}}_{\ \alpha \beta }^{\gamma }$ by introducing d--connection
coefficients (\ref{candcon}) into (\ref{dtorsc})).

By direct computations, we can check that for any given d--connection $\
_{\circ }\Gamma _{\ \beta \gamma }^{\alpha }=\left( \ _{\circ }L_{jk}^{i},\
_{\circ }C_{bc}^{a}\right) $ and nonsymmetric d--metric $\mathbf{\check{g}}=%
\mathbf{g}+\mathbf{a}$ on $\mathbf{V}$ the d--connection $\ _{\ast }\Gamma
_{\ \beta \gamma }^{\alpha }=\left( \ _{\ast }L_{jk}^{i},\ _{\ast
}C_{bc}^{a}\right) ,$ where%
\begin{eqnarray}
\ _{\ast }L_{jk}^{i} &=&\ _{\circ }L_{jk}^{i}+\frac{1}{2}[g^{ir}\ _{\circ
}D_{k}g_{rj}+  \label{aux11} \\
&&\ ^{\pm }O_{sj}^{ir}(\check{a}^{st}\ _{\circ }D_{k}a_{tr}+3\mathit{l}%
_{t}^{s}\ _{\circ }D_{k}\mathit{l}_{r}^{t}-\ _{\circ }D_{k}\mathit{l}%
_{r}^{s})]  \notag \\
\ _{\ast }C_{bc}^{a} &=&\ _{\circ }C_{bc}^{a}+\frac{1}{2}[g^{ah}\ _{\circ
}D_{c}g_{hb}+  \notag \\
&&\ ^{\pm }O_{eb}^{ah}(\check{a}^{ed}\ _{\circ }D_{c}a_{dh}+3\mathit{l}%
_{d}^{e}\ _{\circ }D_{c}\mathit{l}_{h}^{d}-\ _{\circ }D_{c}\mathit{l}%
_{h}^{e})]  \notag
\end{eqnarray}%
is d--metric compatible, i.e satisfies the conditions $\ _{\ast }\mathbf{D%
\check{g}=0.}$

The set of d--connections $\mathbf{D}=\ _{\circ }\mathbf{D+B}$ being
generated by deformations of an arbitrary fixed d--connection $\ _{\circ }%
\mathbf{D}$ in order to be compatible with a given nonsymmetric d--metric $%
\mathbf{\check{g}}=\mathbf{g}+\mathbf{a}$ on $\mathbf{V}$ is defined by
distorsion d--tensors $\mathbf{B=}\left( \ _{h}\mathbf{B,}\ _{v}\mathbf{B}%
\right) $ which can be computed in explicit form, see Ref. \cite{avnsm02}.
In this paper, for simplicity, we shall work with a general d--connection $%
\mathbf{D}$ which is compatible to $\mathbf{\check{g},}$ i.e. satisfies the
conditions (\ref{nscompg}), or (\ref{nscomp}), and can be generated by a
distorsion tensor $\mathbf{B}$ from $\widehat{\mathbf{D}}$ (\ref{candcon}),
or from $_{n}\mathbf{D}$ (\ref{acscomp}). We note for certain canonical
constructions the d--objects $\mathbf{D,}\ _{\circ }\mathbf{D,\widehat{%
\mathbf{D}},\ }_{n}\mathbf{D}$ and $\mathbf{B}$ are completely defined by
the coefficients of a d--metric $\mathbf{\check{g}}=\mathbf{g}+\mathbf{a}$
and $\mathbf{N}$ on $\mathbf{V}.$

Finally, it should be emphasized that because $\ _{\circ }\Gamma _{\ \beta
\gamma }^{\alpha }=\left( \ _{\circ }L_{jk}^{i},\ _{\circ }C_{bc}^{a}\right)
$ is an arbitrary d--connection, it can be chosen to be an important one for
certain physical or geometrical problems. In this work, we shall consider
certain exact solutions in gravity with nonholonomic variables defining a
corresponding $_{\circ }\Gamma _{\ \beta \gamma }^{\alpha }$ and then
deformed to nonsymmetric configurations following formulas (\ref{aux11}).

\subsection{General NGT models with d--connections}

The goal of this section is to analyze N--adapted nonholonomic NGT models\
completely defined by a N--connection $\mathbf{N}=\{N_{i}^{a}\},$ d--metric $%
\mathbf{\check{g}=g+a}$ (\ref{hvm}) and a metric compatible d--connection $%
\Gamma _{\mu \nu }^{\lambda }.$

We follow a N--adapted variational calculus, when instead of partial
derivatives there are used the ''N--elongated'' partial derivatives $\mathbf{%
e}_{\rho }$ (\ref{dder}), varying independently the d--fields $\mathbf{%
\check{g}=g+a}$ and $\ \mathbf{\Gamma }_{\ \beta \gamma }^{\alpha }.$ In
this case, $\check{a}=\left( \check{a}^{ij}\right) $ does not depend on the
choice of fields $\widehat{\xi },$ see (\ref{aux00}), and we can write $%
\mathbf{\check{g}}^{[\rho \sigma ]}=\mathbf{\check{a}}^{\rho \sigma }=[%
\check{a}^{ij},\check{a}^{cb}],$ where $\check{a}^{ij}=-\check{a}^{ji}$ and $%
\check{a}^{cb}=-\check{a}^{bc}.$ We shall work with d--connections,
\begin{equation}
\ \mathbf{W}_{\mu \nu }^{\lambda }\doteqdot \ \mathbf{\Gamma }_{\mu \nu
}^{\lambda }-\frac{2}{3}\delta _{\mu }^{\lambda }\ \mathbf{W}_{\nu },
\label{stbfc}
\end{equation}%
where $\mathbf{W}_{\nu }=\frac{1}{2}\left( \mathbf{W}_{\mu \lambda
}^{\lambda }-\mathbf{W}_{\lambda \mu }^{\lambda }\right) ,$ which means that
$\mathbf{\Gamma }_{\ [\beta \gamma ]}^{\alpha }=0.$ \ This defines a
covariant derivative of type
\begin{equation*}
\ _{W}\mathbf{D}_{\gamma }\mathbf{\check{g}}_{\alpha \beta }=\mathbf{e}%
_{\gamma }\mathbf{\check{g}}_{\alpha \beta }-\mathbf{W}_{\ \alpha \gamma
}^{\tau }\mathbf{\check{g}}_{\tau \beta }-\mathbf{W}_{\ \beta \gamma }^{\tau
}\mathbf{\check{g}}_{\alpha \tau }.
\end{equation*}%
We also can compute
\begin{equation*}
\mathbf{P}_{\mu \nu }\doteqdot \ _{W}\mathbf{R}_{\ \lambda \mu \nu
}^{\lambda }=\mathbf{e}_{\mu }\mathbf{W}_{\ \lambda \nu }^{\lambda }-\mathbf{%
e}_{\nu }\mathbf{W}_{\ \lambda \mu }^{\lambda },
\end{equation*}%
where $\ _{W}\mathbf{R}_{\ \lambda \mu \nu }^{\lambda }$ is computed
following formulas (\ref{dcurvc}) with d--connection $\mathbf{W}$ instead of
$\mathbf{\Gamma .}$ The corresponding to $\mathbf{W}$ and $\ \mathbf{\Gamma }
$ Ricci d--tensors, are related by formulas
\begin{equation*}
\ _{W}\mathbf{R}_{\mu \nu }=\ _{\Gamma }\mathbf{R}_{\mu \nu }+\frac{2}{3}%
\mathbf{e}_{[\nu }\ \mathbf{W}_{\mu ]},
\end{equation*}%
where $\ _{\Gamma }\mathbf{R}_{\mu \nu }$ are given by formulas (\ref{dricci}%
). The variables of this generalized theory, with gravitational constant $%
\left( 16\pi G_{N}\right) ^{-1}=1,$ are parametrized:
\begin{eqnarray*}
\mathbf{\check{g}}_{\mu \nu } &=&\mathbf{g}_{\mu \nu }+\mathbf{a}_{\mu \nu
}+...\mathbf{,}\ \mbox{\qquad  full, nonsymmetric d--metric; } \\
\mathbf{\check{g}}_{(\mu \nu )} &=&\frac{1}{2}(\mathbf{\check{g}}_{\mu \nu }+%
\mathbf{\check{g}}_{\nu \mu })\approx \mathbf{g}_{\mu \nu },\
\mbox{\qquad
symmetric d--metric; } \\
\mathbf{\check{g}}_{[\mu \nu ]} &=&\frac{1}{2}(\mathbf{\check{g}}_{\mu \nu }-%
\mathbf{\check{g}}_{\nu \mu })\approx \mathbf{a}_{\mu \nu },\
\mbox{\qquad
antisymmetric d--metric; } \\
\mathbf{\check{g}}_{\mu \alpha }\mathbf{\check{g}}^{\mu \beta } &=&\mathbf{%
\check{g}}_{\alpha \mu }\mathbf{\check{g}}^{\beta \mu }=\delta _{\alpha
}^{\beta }\neq \mathbf{\check{g}}_{\alpha \mu }\mathbf{\check{g}}^{\mu \beta
}; \\
\mathbf{W}_{\ \beta \gamma }^{\alpha } &\doteqdot &\mathbf{\Gamma }_{\ \beta
\gamma }^{\alpha }-\frac{2}{3}\delta _{\beta }^{\alpha }\mathbf{W}_{\ \gamma
},\mbox{\qquad  full, nonsymmetric d--connection; } \\
\mathbf{W}_{\ \beta } &\doteqdot &\mathbf{W}_{\ [\beta \alpha ]}^{\alpha }.
\end{eqnarray*}

We shall use a nonholonomic generalization of the Lagrangian from \cite%
{ddmcc},%
\begin{eqnarray}
\mathcal{L} &=&\sqrt{-\mathbf{\check{g}}}\mathbf{\check{g}}^{\mu \nu }[\ _{W}%
\mathbf{R}_{\mu \nu }+a_{1}\mathbf{P}_{\mu \nu }+a_{2}\mathbf{e}_{[\mu }\
\mathbf{W}_{\nu ]}+b_{1}\ _{W}\mathbf{D}_{\gamma }\mathbf{W}_{\ [\mu \nu
]}^{\gamma }  \label{genlag} \\
&+& b_{2}\mathbf{W}_{\ [\mu \alpha ]}^{\lambda }\mathbf{W}_{\ [\lambda \nu
]}^{\alpha }+b_{3}\mathbf{W}_{\ [\mu \nu ]}^{\lambda }\mathbf{W}_{\lambda }+%
\mathbf{\check{g}}^{\lambda \delta }\mathbf{\check{g}}_{\alpha \beta }(c_{1}%
\mathbf{W}_{\ [\mu \lambda ]}^{\alpha }\mathbf{W}_{\ [\nu \delta ]}^{\beta }
\notag \\
&+& c_{2}\mathbf{W}_{\ [\mu \nu ]}^{\alpha }\mathbf{W}_{\ [\lambda \delta
]}^{\beta }+c_{3}\mathbf{W}_{\ [\mu \delta ]}^{\alpha }\mathbf{W}_{\ [\nu
\lambda ]}^{\beta }+d_{1}\mathbf{W}_{\mu }\mathbf{W}_{\nu }+2\Lambda )],
\notag
\end{eqnarray}%
where the parameters $a_{1},a_{2},$ etc. are certain constants and $\Lambda $
is the cosmological constant. One should fix certain values of such
constants and take $\mathbf{W}_{\ \beta \gamma }^{\alpha }$ to be defined by
a general affine (in particular, Levi--Civita) connection, in order to get
different Moffat or other models of NGT.

\section{Linearization to Symmetric Anholonomic Backgrounds}

In this Section, we shall prove that for general nonsymmetric metrics
defined on nonholonomic manifolds and corresponding nonholonomic
deformations and linearization a class of general Lagrangians for NGT can be
transformed into stable Lagrangians similar to those used in $\sigma $%
--model and anholonomic and/or noncommutative corrections to general
relativity. We follow the a geometric formalism elaborated in \cite%
{avnsm02,vrfg,vsgg} and reconsider the results of works \cite{jp1,jp2} for
nonholonomic spaces enabled both with nonlinear connection and nonsymmetric
metric structures.

Let us consider an expansion of the Lagrangian (\ref{genlag}) for $\mathbf{%
\check{g}=g+a}$ around a background spacetime defined by a symmetric
d--connection $\mathbf{g}=\{\mathbf{g}_{\alpha \beta }\}$ and a metric
compatible d--connection $\ ^{b}\mathbf{\Gamma }_{\ \beta \gamma }^{\alpha }$
defined by $\mathbf{N}$ and $\mathbf{g}$ (it can be a normal, canonical
d--connection, Cartan or another one) and denote $\mathbf{\check{g}}%
_{[\alpha \beta ]}=\mathbf{a}_{\alpha \beta }.$ We use decompositions of
type
\begin{eqnarray}
\mathbf{\check{g}}_{\alpha \beta } &=&\mathbf{g}_{\alpha \beta }+\ ^{1}%
\mathbf{g}_{\alpha \beta }+...,\ \mathbf{a}_{\alpha \beta }=\ ^{1}\mathbf{a}%
_{\alpha \beta }+\ ^{2}\mathbf{a}_{\alpha \beta }...,  \label{dec1} \\
\mathbf{\Gamma }_{\ \beta \gamma }^{\alpha } &=&\ ^{b}\mathbf{\Gamma }_{\
\beta \gamma }^{\alpha }+\ ^{1}\mathbf{\Gamma }_{\ \beta \gamma }^{\alpha
}+...,\ \ \mathbf{W}_{\mu }=\ ^{1}\mathbf{W}_{\mu }+\ ^{2}\mathbf{W}_{\mu
}+...  \notag
\end{eqnarray}%
when, re--defining $\ ^{1}\mathbf{a}_{\alpha \beta }\rightarrow \ \mathbf{a}%
_{\alpha \beta },$ $\ ^{2}\mathbf{a}_{\alpha \beta }\sim \ \mathbf{a}%
_{..}\cdot \mathbf{a}_{..},$ one holds
\begin{eqnarray}
\mathbf{\check{g}}_{\mu \nu } &=&\mathbf{g}_{\mu \nu }+\mathbf{a}_{\mu \nu
}+\rho \mathbf{a}_{\mu \alpha }\mathbf{a}_{\ \nu }^{\alpha }+\sigma \mathbf{a%
}^{2}\mathbf{g}_{\mu \nu }+O(\mathbf{a}^{3}),  \label{decompm} \\
\mathbf{\check{g}}^{\mu \nu } &=&\mathbf{g}^{\mu \nu }+\mathbf{a}^{\mu \nu
}+(1-\rho )\mathbf{a}^{\mu \alpha }\mathbf{a}_{\alpha }^{\ \nu }+\sigma
\mathbf{a}^{2}\mathbf{g}_{\mu \nu }+O(\mathbf{a}^{3}),  \notag
\end{eqnarray}%
which implies that
\begin{equation*}
\sqrt{|\mathbf{\check{g}}_{\mu \nu }|}=\sqrt{|\mathbf{g}_{\mu \nu }|}\left(
1+\frac{1}{2}(\frac{1}{2}-\rho +4\sigma )\mathbf{a}^{2}\right) ,
\end{equation*}%
for $\mathbf{a}^{2}=\mathbf{a}_{\mu \alpha }\mathbf{a}^{\mu \alpha },$ where
$\mathbf{g}_{\mu \nu }$ and its inverse $\mathbf{g}^{\mu \nu }$ are used to
raise and lower indices. Following a N--adapted calculus with
''N--elongated'' partial differential and differential operators (see (\ref%
{dder}) and (\ref{ddif})) instead of usual partial derivatives and local
coordinate (co) bases, similarly to constructions in Appendix to Ref. \cite%
{avnsm02}, we get from (\ref{genlag}) (up to the second order approximations
on $\mathbf{a})$ the effective Lagrangian
\begin{eqnarray}
\mathcal{L} &=&\sqrt{-g}[\ ^{s}R+2\Lambda -\frac{1}{12}\mathbf{H}^{2}+\left(
\frac{1}{4}\mu ^{2}+\beta \ ^{s}R\right) \mathbf{a}^{2}  \label{eflag} \\
&&-\alpha \mathbf{R}_{\mu \nu }\mathbf{a}^{\mu \alpha }\mathbf{a}_{\alpha
}^{\ \nu }-\gamma \mathbf{R}_{\mu \alpha \nu \beta }\mathbf{a}^{\mu \nu }%
\mathbf{a}^{\alpha \beta }]+O(\mathbf{a}^{3}),  \notag
\end{eqnarray}%
where the effective gauge field (absolutely symmetric torsion) is
\begin{equation}
\mathbf{H}_{\alpha \beta \gamma }=\mathbf{e}_{\alpha }\mathbf{a}_{\beta
\gamma }+\mathbf{e}_{\beta }\mathbf{a}_{\gamma \alpha }+\mathbf{e}_{\gamma }%
\mathbf{a}_{\alpha \beta },  \label{anst}
\end{equation}%
with an effective mass for $\mathbf{a}_{\beta \gamma },$ $\mu ^{2}=2\Lambda
(1-2\rho +8\sigma ),$when the curvature d--tensor $\mathbf{R}_{\mu \alpha
\nu \beta },$ Ricci d--tensor $\mathbf{R}_{\mu \nu }$ and scalar curvature $%
\ ^{s}R$ are correspondingly computed following formulas (\ref{dcurvc}), (%
\ref{dricci}) and (\ref{scalc}), and the constants from (\ref{genlag}) and (%
\ref{decompm}) are re--defined following formulas (\ref{redefc}) in Appendix.

If in the effective Lagrangian (\ref{eflag}) we take instead of a metric
compatible d--connection$\ \mathbf{\Gamma }_{\ \beta \gamma }^{\alpha }$ the
Levi--Civita connection$\ _{\shortmid }\Gamma _{\ \beta \gamma }^{\alpha },$
we get the formula (A29) in \cite{jp1} for nonsymmetric gravitational
interactions modelled on a (pseudo) Riemannian background. It exists a
theorem proven by van Nieuwenhuizen \cite{nieu} stating that in flat space
the only consistent action for a massive antisymmetric tensor field is of
the form
\begin{equation}
\ ^{fl}\mathcal{L}=-\frac{1}{12}\mathbf{H}^{2}+\frac{1}{4}\mu ^{2}\mathbf{a}%
^{2}+O(\mathbf{a}^{3}),  \label{gfspel}
\end{equation}%
for $\mathbf{a}^{2}=\mathbf{a}^{\mu \nu }\mathbf{a}_{\mu \nu }.$ A rigorous
study provided in \cite{jp1} proves that $\gamma =0,$ see (\ref{eflag}), is
not allowed in NGT extended nearly a Schwarzschild background because in
such a case it is not possible to solve in a compatible form the conditions (%
\ref{aux4}) for $\gamma =\Xi =0.$

A quite general solution of the problem of instability in NGT found by
Janssen and Prokopec is to compensate the term with $\gamma \neq 0$ in (\ref%
{eflag}). To do this, we can constrain such a way the nonholonomic frame
dynamics \footnote{%
in explicit form, we have to impose certain constraints on coefficients $%
N_{i}^{a}$ from (\ref{ansatz}) and (\ref{shvm}), see the end of this Section}
when we get for decompositions of NGT\ with respect to any general
relativity background an effective Lagrangian without coupling of spacetime
curvature tensors with nonsymmetric tensor $\mathbf{b}^{\mu \alpha }$ (i.e.
without a term of type $\gamma \ _{\shortmid }R_{\mu \alpha \nu \beta }%
\mathbf{b}^{\mu \nu }\mathbf{b}^{\alpha \beta }),$
\begin{equation}
\ ^{E}\mathcal{L}=\sqrt{-g}[\ _{\shortmid }R+2\ _{\shortmid }\Lambda -\frac{1%
}{12}\ \mathbf{H}^{2}+\left( \frac{1}{4}\mu ^{2}+\beta \ \ _{\shortmid
}R\right) \mathbf{b}^{2}-\alpha \ _{\shortmid }R_{\mu \nu }\mathbf{b}^{\mu
\alpha }\mathbf{b}_{\alpha }^{\ \nu }]+O(b^{3}).  \label{goodlagr}
\end{equation}%
In this formula $\ _{\shortmid }R$ and $\ _{\shortmid }R_{\mu \nu }$ are
respectively the scalar curvature and the Ricci tensor computed for $\
_{\shortmid }\mathbf{\Gamma }_{\ \beta \gamma }^{\alpha },$ see formulas (%
\ref{aux01}) in Appendix and $\ _{\shortmid }\Lambda $ is an effective
cosmological constant with possible small polarizations depending on $%
u^{\alpha }.$

We show how for a class of nonholonomic deformations of general relativity
backgrounds, we get effective Lagrangians which seem to have a good flat
spacetime limit of type (\ref{gfspel}):

Let us consider $N_{i}^{a}\approx \mathring{\varepsilon}^{2}$ $n_{i}^{a}$
and $\mathbf{a}^{\mu \alpha }\approx \mathring{\varepsilon}$ $\mathbf{b}%
^{\mu \alpha }$ and take $\ ^{b}\mathbf{\Gamma }_{\ \beta \gamma }^{\alpha }=%
\widehat{\mathbf{\Gamma }}_{\ \alpha \beta }^{\gamma }$ (\ref{candcon}) in
decomposition for d--connection (\ref{dec1}), where $\mathring{\varepsilon}$
is a small parameter, which results (following formulas (\ref{cdeft}), (\ref%
{anst}), (\ref{dder}) and (\ref{aux01})) in deformations of type
\begin{eqnarray}
\ _{\shortmid }\Gamma _{\ \alpha \beta }^{\gamma } &=&\widehat{\mathbf{%
\Gamma }}_{\ \alpha \beta }^{\gamma }+\mathring{\varepsilon}^{2}\
_{\shortmid }\mathring{z}_{\ \alpha \beta }^{\gamma }(n_{i}^{a})...,
\label{aux02s} \\
\ ^{s}R &=&\ _{\shortmid }R+\mathring{\varepsilon}^{2}\ _{\shortmid }%
\mathring{z}(n_{i}^{a})...,\mathbf{H}^{2}=\mathring{\varepsilon}^{2}\
_{\shortmid }H(\mathbf{b}^{\mu \alpha }),  \notag
\end{eqnarray}%
where $\ _{\shortmid }H(\mathbf{b}^{\mu \alpha })$ is computed by formula (%
\ref{anst}) with $\mathbf{e}_{\alpha }\rightarrow \partial _{\alpha }$ and $%
\mathbf{a}_{\beta \gamma }\rightarrow \mathbf{b}_{\beta \gamma }$ and the
functionals $\ _{\shortmid }z_{\ \alpha \beta }^{\gamma
}(g_{ij,}g_{ab},n_{i}^{a})$ and $\ \ _{\shortmid }\mathring{z}%
(g_{ij,}g_{ab},n_{i}^{a})$ can be computed by introducing (\ref{deft}) into
respective formulas for connections and scalar curvature. Introducing values
(\ref{aux02s}) into (\ref{gfspel}) and identifying
\begin{equation}
\ _{\shortmid }\Lambda \approx \Lambda ,  \label{efpolc}
\end{equation}%
we get that $\mathcal{L}\rightarrow \ ^{E}\mathcal{L}$ if and only if%
\begin{equation}
\ _{\shortmid }z(g_{ij,}g_{ab},\mathring{n}_{i}^{a})=\gamma \ _{\shortmid
}R_{\mu \alpha \nu \beta }\mathbf{b}^{\mu \nu }\mathbf{b}^{\alpha \beta }.
\label{deftsc}
\end{equation}%
The left part of this equation is defined by the quadratic $\mathring{%
\varepsilon}^{2}$ deformation of scalar curvature, from $\ _{\shortmid }R$
to $\ ^{s}R,$ relating algebraically the coefficients $g_{ij},h_{ab}$ and $%
n_{i}^{a}$ and their partial derivatives. We do not provide in this work the
cumbersome formula for $\ _{\shortmid }\mathring{z}(g_{ij,}g_{ab},n_{i}^{a})$
in the case of general nonholonomic or Einstein gravity backgrounds, but we
shall compute it explicitly and solve the equation (\ref{deftsc}) for an
ellipsoidal background in next section. Here we emphasize that in theories
with zero cosmological constant we have to consider $\ _{\shortmid }\Lambda
\approx \Lambda =0.$

We conclude that we are able to generate stable NGT gravity models on
backgrounds with small nonholonomic frame and nonsymmetric metric
deformations if the conditions (\ref{deftsc}) are satisfied. This induces a
small locally anisotropic polarization of the cosmological constant, see (%
\ref{efpolc}). Having stabilized the gravitational interactions with the
nonsymmetric components of metric, for certain gravitational configurations
with another small parameter $\varepsilon \rightarrow 0,$ we get certain
backgrounds in general relativity (for instance, the Schwarzschild one). For
generic nonlinear theories, such as NGT and the Einstein gravity, the
procedures of constraining certain nonlinear solutions in order to get
stable configurations and taking smooth limits on a small parameter
resulting in holonomic backgrounds are not commutative.

Finally, we note that we can use similar decompositions of type (\ref{aux02s}%
) to transform an arbitrary metric compatible d--connection $\mathbf{\Gamma }%
_{\ \alpha \beta }^{\gamma }$ to $\widehat{\mathbf{\Gamma }}_{\ \alpha \beta
}^{\gamma },$ and/or to introduce two small parameters consider deformations
of type $\mathbf{\Gamma }_{\ \alpha \beta }^{\gamma }\rightarrow $ $\widehat{%
\mathbf{\Gamma }}_{\ \alpha \beta }^{\gamma }\rightarrow \ _{\shortmid
}\Gamma _{\ \alpha \beta }^{\gamma }.$ We shall use this approach in the
next section.

\section{Stability of Stationary Ellipsoidal Solutions}

The effective gravitational field equations for nonsymmetric metrics on
symmetric nonholonomic backgrounds are derived. We also analyze a class of
solutions in NGT on a nonholonomic ellipsoidal background. For vanishing
eccentricity, such solutions have nontrivial limits to Schwarzschild
configurations.

\subsection{Field equations with nonholonomic backgrounds}

The field equations derived from an effective Lagrangian (\ref{eflag}) for a
d--connection $\mathbf{\Gamma }_{\ \alpha \beta }^{\gamma }$ are
\begin{eqnarray*}
\left( \sqrt{|\mathbf{g}_{\mu \nu }|}\right) ^{-1}\mathbf{e}_{\alpha }(\sqrt{%
|\mathbf{g}_{\mu \nu }|}\mathbf{H}^{\alpha \beta \nu })+(\mu ^{2}+4\beta \
^{s}R)\mathbf{a}^{\beta \nu } && \\
+4\alpha \mathbf{a}^{\alpha (\nu }\mathbf{R}_{\ \alpha }^{\beta )}+4\gamma
\mathbf{a}^{\alpha \tau }\mathbf{R}_{\ \alpha \tau }^{\beta \quad \nu }+%
\mathcal{O}(\mathbf{a}^{2}) &=&0, \\
\mathbf{R}_{\mu \nu }-\frac{1}{2}\mathbf{g}_{\mu \nu }\ ^{s}R-\Lambda
\mathbf{g}_{\mu \nu }+\mathcal{O}(\mathbf{a}^{2}) &=&0.
\end{eqnarray*}%
We shall work with two--parameter, deformations of nonlinear and linear
connections, respectively of%
\begin{equation*}
N_{i}^{a}\approx \varepsilon n_{i}^{a}+\mathring{\varepsilon}^{2}\mathring{n}%
_{i}^{a}+...\mbox{ \and }\mathbf{a}^{\mu \alpha }\approx \mathring{%
\varepsilon}\mathbf{b}^{\mu \alpha }
\end{equation*}%
and
\begin{eqnarray*}
\widehat{\mathbf{\Gamma }}_{\ \alpha \beta }^{\gamma } &=&\mathbf{\Gamma }%
_{\ \alpha \beta }^{\gamma }+\mathring{\varepsilon}^{2}\ \mathring{z}_{\
\alpha \beta }^{\gamma }(\mathring{n}_{i}^{a})+... \\
\ _{\shortmid }\Gamma _{\ \alpha \beta }^{\gamma } &=&\widehat{\mathbf{%
\Gamma }}_{\ \alpha \beta }^{\gamma }+\varepsilon \ _{\shortmid }z_{\ \alpha
\beta }^{\gamma }(n_{i}^{a})...,
\end{eqnarray*}%
for
\begin{eqnarray*}
\ ^{s}R &=&\ ^{s}\widehat{R}+\mathring{\varepsilon}^{2}\ \mathring{z}%
(g_{ij,}g_{ab},n_{i}^{a})...,\mathbf{H}^{2}=\mathring{\varepsilon}^{2}%
\mathbf{\mathring{H}}(\mathbf{b}^{\mu \alpha }), \\
\ ^{s}\widehat{R} &=&\ _{\shortmid }R+\varepsilon \ _{\shortmid
}z(g_{ij,}g_{ab},n_{i}^{a})+...,
\end{eqnarray*}%
where
\begin{equation}
\Lambda \approx \ \mathring{\varepsilon}^{2}\mathring{\Lambda},
\label{const1}
\end{equation}%
we transform $\mathcal{L}$ (\ref{eflag}) into
\begin{equation}
\mathcal{\mathring{L}=}\sqrt{-g}[\ ^{s}\widehat{R}+2\mathring{\Lambda}-\frac{%
\ \mathbf{\mathring{H}}^{2}}{12}+\left( \frac{\mu ^{2}}{4}+\beta \ ^{s}%
\widehat{R}\right) \mathbf{b}^{2}-\alpha \ \widehat{\mathbf{R}}_{\mu \nu }%
\mathbf{b}^{\mu \alpha }\mathbf{b}_{\alpha }^{\ \nu }]+\mathcal{O}(\mathbf{b}%
^{3})  \label{ellipl}
\end{equation}%
if and only if%
\begin{equation}
\mathring{z}(g_{ij,}g_{ab},n_{i}^{a})=\gamma \ \widehat{\mathbf{R}}_{\mu
\alpha \nu \beta }\mathbf{b}^{\mu \nu }\mathbf{b}^{\alpha \beta }.
\label{stabcond}
\end{equation}

The N--adapted variational field equations derived from (\ref{ellipl}) are
\begin{eqnarray}
\frac{\mathbf{e}_{\alpha }(\sqrt{|\mathbf{g}_{\mu \nu }|}\mathbf{\mathring{H}%
}^{\alpha \beta \nu })}{\sqrt{|\mathbf{g}_{\mu \nu }|}}+(\mu ^{2}+4\beta \
^{s}\widehat{R})\mathbf{b}^{\beta \nu }+4\alpha \mathbf{b}^{\alpha (\nu }%
\widehat{\mathbf{R}}_{\ \alpha }^{\beta )}+\mathcal{O}(\mathbf{b}^{2}) &=&0,
\label{feq1} \\
\widehat{\mathbf{R}}_{\mu \nu }-\frac{1}{2}\mathbf{g}_{\mu \nu }\ ^{s}%
\widehat{R}-\ \mathring{\varepsilon}^{2}\mathring{\Lambda}\mathbf{g}_{\mu
\nu }+\mathcal{O}(\mathbf{b}^{2}) &=&0,  \label{feq2}
\end{eqnarray}%
where $\widehat{\mathbf{R}}_{\ \alpha \tau \nu }^{\beta \quad },\widehat{%
\mathbf{R}}_{\mu \nu }$ and $\ ^{s}\widehat{R}$ are computed respectively by
introducing the coefficients (\ref{candcon}) into formulas (\ref{dcurvc}), (%
\ref{dricci}) and (\ref{scalc}). We can see that to the order $\mathcal{O}(%
\mathbf{b}^{2})$ the fields equations decouple on the symmetric and
nonsymmetric parts of d--metrics which allows us to consider a nonholonomic
symmetric background defined by $(\mathbf{g}_{\mu \nu },N_{i}^{a},\widehat{%
\mathbf{\Gamma }}_{\ \alpha \beta }^{\gamma })$ and to reduce the problem to
the study of constrained dynamics of the antisymmetric d--field $\mathbf{a}%
^{\beta \nu }$ on this background.

\subsection{Solutions with ellipsoidal symmetry}

The simplest class of solutions for the system (\ref{feq1}) and (\ref{feq1})
can be constructed in the approximation that $\mathring{\varepsilon}^{2}%
\mathring{\Lambda}\sim 0$ and $\mu ^{2}\sim 0.$ \footnote{%
As a matter of principle, we can consider solutions with nonzero values of
mass $\mu ,$ but this will result in more sophisticate configurations for
the nonsymmetric components of metrics which is not related to the problem
of nonholonomic stabilization of NGT; see Chapter 3 in Ref. \cite{vsgg}, for
similar details on constructing static black ellipsoid solutions in gravity
with nonholonomic completely antisymmetric metric defined as a Proca field,
and \cite{vncg}, for complex generalizations of such solutions to
noncommutative gravity.} For the ansatz
\begin{equation}
\mathbf{\mathring{H}}_{\alpha \beta \nu }=\ ^{b}\lambda \sqrt{|\mathbf{g}%
_{\mu \nu }|}\epsilon _{\alpha \beta \nu },  \label{anstors}
\end{equation}%
where $\ ^{b}\lambda =const$ and $\epsilon _{\alpha \beta \nu }$ being the
complete antisymmetric tensor, and any (vacuum) solution for
\begin{equation}
\widehat{\mathbf{R}}_{\mu \nu }=0,  \label{vnonheq}
\end{equation}%
we generate decoupled solutions both for the symmetric and nonsymmetric part
of metric. The nonsymmetric field $\mathbf{b}_{\beta \gamma }$ is any
solution of
\begin{equation}
\ ^{b}\lambda \sqrt{|\mathbf{g}_{\mu \nu }|}\epsilon _{\alpha \beta \nu }=%
\mathbf{e}_{\alpha }\mathbf{b}_{\beta \gamma }+\mathbf{e}_{\beta }\mathbf{b}%
_{\gamma \alpha }+\mathbf{e}_{\gamma }\mathbf{b}_{\alpha \beta },
\label{ansymtors}
\end{equation}%
which follows from formulas (\ref{anst}) and (\ref{anstors}).

\subsubsection{Anholonomic deformations of the Schwarzschild metric}

Let us consider a primary quadratic element
\begin{equation}
\delta s_{[1]}^{2}=-d\xi ^{2}-r^{2}(\xi )\ d\vartheta ^{2}-r^{2}(\xi )\sin
^{2}\vartheta \ d\varphi ^{2}+\varpi ^{2}(\xi )\ dt^{2},  \label{5aux1}
\end{equation}%
where the local coordinates and nontrivial metric coefficients are
parametriz\-ed in the form%
\begin{eqnarray}
x^{1} &=&\xi ,x^{2}=\vartheta ,y^{3}=\varphi ,y^{4}=t,  \label{5aux1p} \\
\check{g}_{1} &=&-1,\ \check{g}_{2}=-r^{2}(\xi ),\ \check{h}_{3}=-r^{2}(\xi
)\sin ^{2}\vartheta ,\ \check{h}_{4}=\varpi ^{2}(\xi ),  \notag
\end{eqnarray}%
for
\begin{equation*}
\xi =\int dr\ \left| 1-\frac{2m_{0}}{r}+\frac{\varepsilon }{r^{2}}\right|
^{1/2}\mbox{\ and\ }\varpi ^{2}(r)=1-\frac{2m_{0}}{r}+\frac{\varepsilon }{%
r^{2}}.
\end{equation*}%
For the constants $\varepsilon \rightarrow 0$ and $m_{0}$ being a point
mass, the element (\ref{5aux1}) defines the Schwarzschild solution written
in spacetime spherical coordinates $(r,\vartheta ,\varphi ,t).$ The
parameter $\varepsilon $ should not be confused with the square of the
electric charge $e^{2}$ for the Reissner--Nordstr\"{o}m metric. In our
further considerations, we treat $\varepsilon $ as a small parameter, for
instance, defining a small deformation of a circle into an ellipse
(eccentricity).

We construct a generic off--diagonal vacuum solution\footnote{%
it can not diagonalized by coordinate transforms} by using nonholonomic
deformations, $g_{i}=\eta _{i}\check{g}_{i}$ and $h_{a}=\eta _{a}\check{h}%
_{a},$ where $(\check{g}_{i},\check{h}_{a})$ are given by data (\ref{5aux1p}%
), when the new ansatz (target metric),
\begin{eqnarray}
\delta s_{[def]}^{2} &=&-\eta _{1}(\xi )d\xi ^{2}-\eta _{2}(\xi )r^{2}(\xi
)\ d\vartheta ^{2}  \label{5sol1} \\
&&-\eta _{3}(\xi ,\vartheta ,\varphi )r^{2}(\xi )\sin ^{2}\vartheta \ \delta
\varphi ^{2}+\eta _{4}(\xi ,\vartheta ,\varphi )\varpi ^{2}(\xi )\ \delta
t^{2},  \notag \\
\delta \varphi &=&d\varphi +w_{1}(\xi ,\vartheta ,\varphi )d\xi +w_{2}(\xi
,\vartheta ,\varphi )d\vartheta ,\   \notag \\
\delta t &=&dt+n_{1}(\xi ,\vartheta )d\xi +n_{2}(\xi ,\vartheta )d\vartheta ,
\notag
\end{eqnarray}%
is supposed to solve the equation (\ref{vnonheq}). In formulas (\ref{5sol1})
there are used 3D spacial spherical coordinates, $(\xi (r),\vartheta
,\varphi )$ or $(r,\vartheta ,\varphi ).$ The details on determining certain
classes of coefficients for the target metric solving the vacuum Einstein
equations for the canonical d--connection can be found in Refs. \cite%
{avnsm01,vbe1,vncg,ijgmmp} and Part II in \cite{vsgg}. Here we summarize the
results which can be verified by direct computations:

The functions $\eta _{3}$ and $\eta _{4}$ can be generated by a function $%
b(\xi ,\vartheta ,\varphi )$ following the conditions
\begin{equation*}
-h_{0}^{2}(b^{\ast })^{2}=\eta _{3}(\xi ,\vartheta ,\varphi )r^{2}(\xi )\sin
^{2}\vartheta \mbox{ and }b^{2}=\eta _{4}(\xi ,\vartheta ,\varphi )\varpi
^{2}(\xi ),
\end{equation*}%
for
\begin{equation}
|\eta _{3}|=(h_{0})^{2}|\check{h}_{4}/\check{h}_{3}|\left[ \left( \sqrt{%
|\eta _{4}|}\right) ^{\ast }\right] ^{2},  \label{5eq23a}
\end{equation}%
with $h_{0}=const,$ where $\check{h}_{a}$ are stated by the Schwarzschild
solution for the chosen system of coordinates and $\eta _{4}$ can be any
function satisfying the condition $\eta _{4}^{\ast }=\partial \eta
_{4}/\partial \varphi \neq 0.$ We can compute the polarizations $\eta _{1}$
and $\eta _{2},$ when $\eta _{1}=\eta _{2}r^{2}=e^{\psi (\xi ,\vartheta )}$
with $\psi $ solving
\begin{equation*}
\frac{\partial ^{2}\psi }{\partial \xi ^{2}}+\frac{\partial ^{2}\psi }{%
\partial \vartheta ^{2}}=0.
\end{equation*}%
The nontrivial values of N--connection coefficients $N_{i}^{3}=w_{i}(\xi
,\vartheta ,\varphi )$ and $N_{i}^{4}=n_{i}(\xi ,\vartheta ,\varphi ),$ when
$i=1,2,$ for vacuum configurations with the Levi--Civita connection $\nabla $%
\ are given by
\begin{equation*}
w_{1}=\partial _{\xi }(\sqrt{|\eta _{4}|}\varpi )/\left( \sqrt{|\eta _{4}|}%
\right) ^{\ast }\varpi ,\ w_{2}=\partial _{\vartheta }(\sqrt{|\eta _{4}|}%
)/\left( \sqrt{|\eta _{4}|}\right) ^{\ast }
\end{equation*}%
and any $n_{1,2}=\ ^{1}n_{1,2}(\xi ,\vartheta )$ for which $\partial
_{\vartheta }(\ ^{1}n_{1})-\partial _{\xi }(\ ^{1}n_{2})=0,$ when, for
instance $\partial _{\xi }=\partial /\partial \xi .$ In a more general case,
when $\nabla \neq \widehat{\mathbf{D}},$ but the nonholonomic vacuum
equation (\ref{vnonheq}) is solved, we have to take
\begin{equation*}
n_{1,2}(\xi ,\vartheta ,\varphi )=\ ^{1}n_{1,2}(\xi ,\vartheta )+\
^{2}n_{1,2}(\xi ,\vartheta )\int d\varphi ~h_{3}/\left( \sqrt{\left|
h_{4}\right| }\right) ^{3},
\end{equation*}%
for $\ ^{1}n_{1,2}(\xi ,\vartheta )$ and $\ ^{2}n_{1,2}(\xi ,\vartheta )$
being certain integration functions to be defined from certain boundary
conditions, or constrained additionally to solve certain compatibility
equations in some limits.

Putting the defined values of the coefficients in the ansatz (\ref{5sol1}),
we construct a class of exact vacuum solutions of the Einstein equations for
the canonical d--connection (in particular, for the Levi--Civita connection)
defining stationary nonholonomic deformations of the Schwarzschild metric,
\begin{eqnarray}
\delta s_{[1]}^{2} &=&-e^{\psi }\left( d\xi ^{2}+\ d\vartheta ^{2}\right)
-h_{0}^{2}\left[ \left( \sqrt{|\eta _{4}|}\right) ^{\ast }\right] ^{2}\varpi
^{2}\ \delta \varphi ^{2}+\eta _{4}\varpi ^{2}\ \delta t^{2},  \label{5sol1a}
\\
\delta \varphi &=&d\varphi +\frac{\partial _{\xi }(\sqrt{|\eta _{4}|}\varpi )%
}{\left( \sqrt{|\eta _{4}|}\right) ^{\ast }\varpi }d\xi +\frac{\partial
_{\vartheta }(\sqrt{|\eta _{4}|})}{\left( \sqrt{|\eta _{4}|}\right) ^{\ast }}%
d\vartheta ,\   \notag \\
\delta t &=&dt+n_{1}d\xi +n_{2}d\vartheta .  \notag
\end{eqnarray}%
Such solutions were constructed to define anholonomic transform of a static
black hole solution into stationary vacuum Einstein (non)holonomic spaces
with local anisotropy (on coordinate $\varphi )$ defined by an arbitrary
function $\eta _{4}(\xi ,\vartheta ,\varphi )$ with $\partial _{\varphi
}\eta _{4}\neq 0$, an arbitrary $\psi (\xi ,\vartheta )$ solving the 2D
Laplace equation and certain integration functions $\ ^{1}n_{1,2}(\xi
,\vartheta )$ and integration constant $h_{0}^{2}.$ In general, the
solutions from the target set of metrics do not define black holes and do
not describe obvious physical situations. Nevertheless, they preserve the
singular character of the coefficient $\varpi ^{2}$ vanishing on the horizon
of a Schwarzschild black hole if we take only smooth integration functions.
We can also consider a prescribed physical situation when, for instance, $%
\eta _{4}$ mimics 3D, or 2D, solitonic polarizations on coordinates $\xi
,\vartheta ,\varphi ,$ or on $\xi ,\varphi ,$ see Refs. \cite%
{avnsm01,vncg,ijgmmp}.

\subsubsection{Solutions with small nonholonomic polarizations}

The class of solutions (\ref{asol3}) in defined in a very general form. Let
us extract a subclasses of solutions related to the Schwarzschild metric. We
consider decompositions on a small parameter $0<\varepsilon <1$ in (\ref%
{5sol1a}), when
\begin{eqnarray}
\sqrt{|\eta _{3}|} &=&q_{3}^{\hat{0}}(\xi ,\vartheta ,\varphi )+\varepsilon
q_{3}^{\hat{1}}(\xi ,\vartheta ,\varphi )+\varepsilon ^{2}q_{3}^{\hat{2}%
}(\xi ,\vartheta ,\varphi )...,  \notag \\
\sqrt{|\eta _{4}|} &=&1+\varepsilon q_{4}^{\hat{1}}(\xi ,\vartheta ,\varphi
)+\varepsilon ^{2}q_{4}^{\hat{2}}(\xi ,\vartheta ,\varphi )...,  \notag
\end{eqnarray}%
where the ''hat'' indices label the coefficients multiplied to $\varepsilon
,\varepsilon ^{2},...$ \ The conditions (\ref{5eq23a}) are expressed in the
form
\begin{equation*}
\varepsilon h_{0}\sqrt{|\frac{\check{h}_{4}}{\check{h}_{3}}|\ }\left( q_{4}^{%
\hat{1}}\right) ^{\ast }=q_{3}^{\hat{0}},\ \varepsilon ^{2}h_{0}\sqrt{|\frac{%
\check{h}_{4}}{\check{h}_{3}}|\ }\left( q_{4}^{\hat{2}}\right) ^{\ast
}=\varepsilon q_{3}^{\hat{1}},...
\end{equation*}%
We take the integration constant, for instance, to satisfy the condition $%
\varepsilon h_{0}=1$ (choosing a corresponding distributions and system of
coordinates). This condition will be important in order to get stable
solutions for certain $\varepsilon \neq 0,$ but small, i.e. $0<\varepsilon
<1.$ For such small deformations, we prescribe a function $q_{3}^{\hat{0}}$
and define $q_{4}^{\hat{1}},$ integrating on $\varphi $ (or inversely,
prescribing $q_{4}^{\hat{1}},$ then taking the partial derivative $\partial
_{\varphi },$ to compute $q_{3}^{\hat{0}}).$ In a similar form, there are
related the coefficients $q_{3}^{\hat{1}}$ and $q_{3}^{\hat{2}}.$ An
important physical situation arises when we select the conditions when such
small nonholonomic deformations define rotoid configurations. This is
possible, for instance, if
\begin{equation}
2q_{4}^{\hat{1}}=\frac{q_{0}(r)}{4m_{0}^{2}}\sin (\omega _{0}\varphi
+\varphi _{0})-\frac{1}{r^{2}},  \label{5aux1sd}
\end{equation}%
where $\omega _{0}$ and $\varphi _{0}$ are constants and the function $%
q_{0}(r)$ has to be defined by fixing certain boundary conditions for
polarizations. In this case, the coefficient before $\delta t^{2}$ is
\begin{equation}
\eta _{4}\varpi ^{2}=1-\frac{2m_{0}}{r}+\varepsilon (\frac{1}{r^{2}}+2q_{4}^{%
\hat{1}}).  \label{apr01}
\end{equation}%
This coefficient vanishes and defines a small deformation of the
Schwarz\-schild spherical horizon into a an ellipsoidal one (rotoid
configuration) given by
\begin{equation*}
r_{+}\simeq \frac{2\mu }{1+\varepsilon \frac{q_{0}(r)}{4m_{0}^{2}}\sin
(\omega _{0}\varphi +\varphi _{0})}.
\end{equation*}%
Such solutions with ellipsoid symmetry seem to define static black
ellipsoids which are stable (they were investigated in details in Refs. \cite%
{vbe1,vbe2}). The ellipsoid configurations were proven to be stable under
perturbations and transform into the Schwarzschild solution far away from
the ellipsoidal horizon. In general relativity, this class of vacuum metrics
violates the conditions of black hole uniqueness theorems \cite{heu} because
the ''surface'' gravity is not constant for stationary black ellipsoid
deformations.

We can construct an infinite number of ellipsoidal locally anisotropic black
hole deformations. Nevertheless, they present physical interest because they
preserve the spherical topology, have the Minkowski asymptotic and the
deformations can be associated to certain classes of geometric spacetime
distorsions related to generic off--diagonal metric terms. Putting $\varphi
_{0}=0,$ in the limit $\omega _{0}\rightarrow 0,$ we get $q_{4}^{\hat{1}%
}\rightarrow 0$ in (\ref{5aux1sd}). To get a smooth limit to the
Schwarzschild solution we have to state the limit $q_{3}^{\hat{0}%
}\rightarrow 1$ for $\varepsilon \rightarrow 0.$

Let us summarize the above presented approximations for ellipsoidal
symmetries: For (\ref{apr01}), we have
\begin{equation*}
h_{4}=\eta _{4}(\xi ,\vartheta ,\varphi )\varpi ^{2}(\xi )=1-\frac{2m_{0}}{r}%
+\varepsilon \frac{q_{0}(r)}{4m_{0}}\sin (\omega _{0}\varphi +\varphi _{0})+%
\mathcal{O}(\varepsilon ^{2})
\end{equation*}%
and
\begin{equation*}
h_{3}=\eta _{3}(\xi ,\vartheta ,\varphi )r^{2}(\xi )\sin ^{2}\vartheta
=h_{0}^{2}\left[ (\sqrt{|\eta _{4}|})^{\ast }\right] ^{2}\varpi ^{2}(\xi
)=(\varepsilon h_{0})^{2}\left[ (q_{4}^{\hat{1}})^{\ast }\right] ^{2},
\end{equation*}%
which results in
\begin{equation*}
h_{3}=(\varepsilon h_{0})^{2}\frac{q_{0}(r)\omega _{0}^{2}}{16m_{0}}\cos
^{2}(\omega _{0}\varphi +\varphi _{0})+\mathcal{O}(\varepsilon ^{3}),
\end{equation*}%
where we must preserve the second order on $\varepsilon ^{2}$ if $%
\varepsilon h_{0}\sim 1.$ To get a smooth limit of off--diagonal
coefficients in solutions to the\ Schwarzschild metric (\ref{5aux1}), we
state that after integrations one approximates the N--connection
coefficients as $N_{i}^{a}\sim \varepsilon n_{i}^{a}.$ Putting together all
decompositions of coefficients on $\varepsilon $ in (\ref{5sol1a}), we get a
family of ellipsoidal solution of equations (\ref{vnonheq}) decomposed on
eccentricity $\varepsilon ,$
\begin{eqnarray}
\delta s_{[1]}^{2} &=&-e^{\varepsilon \psi }\left( d\xi ^{2}+\ d\vartheta
^{2}\right) -(\varepsilon h_{0})^{2}\frac{q_{0}(r)\omega _{0}^{2}}{16m_{0}}%
\cos ^{2}(\omega _{0}\varphi +\varphi _{0})\ \delta \varphi ^{2}
\label{soleldec} \\
&&+\left[ 1-\frac{2m_{0}}{r}+\varepsilon \frac{q_{0}(r)}{4m_{0}}\sin (\omega
_{0}\varphi +\varphi _{0})+\mathcal{O}(\varepsilon ^{2})\right] \ \delta
t^{2},  \notag \\
\delta \varphi &=&d\varphi +\varepsilon \frac{\partial _{\xi }(\sqrt{|\eta
_{4}|}\varpi )}{\left( \sqrt{|\eta _{4}|}\right) ^{\ast }\varpi }d\xi
+\varepsilon \frac{\partial _{\vartheta }(\sqrt{|\eta _{4}|})}{\left( \sqrt{%
|\eta _{4}|}\right) ^{\ast }}d\vartheta ,\   \notag \\
\delta t &=&dt+\varepsilon n_{1}d\xi +\varepsilon n_{2}d\vartheta .  \notag
\end{eqnarray}

One can be defined certain more special cases when $q_{4}^{\hat{2}}$ and $%
q_{3}^{\hat{1}}$ (as a consequence) are of solitonic locally anisotropic
nature. In result, such solutions will define small stationary deformations
of the Schwarzschild solution embedded into a background polarized by
anisotropic solitonic waves.

Now, we show how we can solve the problem of stability related to the
condition (\ref{stabcond}): Let us consider a small cosmological constant of
type (\ref{const1}) stated only in the horizontal spacetime distribution $\
^{h}\Lambda \approx \ \mathring{\varepsilon}^{2}\ ^{h}\mathring{\Lambda},$
but$\ ^{v}\Lambda =0.$\footnote{%
The techniques presented in Refs. \cite{avnsm01,vbe1,vsgg,ijgmmp} allows us
to construct solutions for nontrivial values $\ ^{v}\Lambda ,$ but this
would result in modifications of the formulas for the vertical part of
d--metric and N--connection coefficients, which is related to a more
cumbersome calculus; in this work, we analyze the simplest examples.} By
straightforward computations, we can verify that the symmetric part of the
ansatz
\begin{eqnarray}
\delta s_{[1]}^{2} &=&-e^{\varepsilon \psi +\mathring{\varepsilon}^{2}%
\mathring{\psi}}\left( d\xi ^{2}+\ d\vartheta ^{2}\right) -(\varepsilon
h_{0})^{2}\frac{q_{0}(r)\omega _{0}^{2}}{16m_{0}}\cos ^{2}(\omega
_{0}\varphi +\varphi _{0})\ \delta \varphi ^{2} +  \notag \\
&&\left[ 1-\frac{2m_{0}}{r}+\varepsilon \frac{q_{0}(r)}{4m_{0}}\sin (\omega
_{0}\varphi +\varphi _{0})+\mathcal{O}(\varepsilon ^{2})\right] \ \delta
t^{2}+\mathring{\varepsilon}\mathbf{b}_{\alpha \beta }\ \mathbf{e}^{\alpha
}\wedge \mathbf{e}^{\beta },  \notag \\
\delta \varphi &=&d\varphi +\varepsilon \frac{\partial _{\xi }(\sqrt{|\eta
_{4}|}\varpi )}{\left( \sqrt{|\eta _{4}|}\right) ^{\ast }\varpi }d\xi
+\varepsilon \frac{\partial _{\vartheta }(\sqrt{|\eta _{4}|})}{\left( \sqrt{%
|\eta _{4}|}\right) ^{\ast }}d\vartheta ,\   \label{asol3} \\
\delta t &=&dt+\varepsilon n_{1}d\xi +\varepsilon n_{2}d\vartheta .  \notag
\end{eqnarray}%
solves the equations
\begin{eqnarray}
R_{ij} &=&\widehat{R}_{ij}+\mathring{\varepsilon}^{2}\mathring{z}_{ij},%
\mbox{ \ for }\mathring{z}_{ij}=-\ ^{h}\mathring{\Lambda}e^{\varepsilon \psi
+\mathring{\varepsilon}^{2}\mathring{\psi}}\delta _{ij},\widehat{R}_{ij}=0
\label{aux3a} \\
R_{ia} &=&\widehat{R}_{ia}=0,\ R_{ai}=\widehat{R}_{ai}=0,\ R_{ab}=\widehat{R}%
_{ab}=0,  \notag
\end{eqnarray}%
see formulas for $\mathbf{\mathbf{R}_{\beta \gamma }}$ (\ref{dricci}),where $%
\mathbf{e}^{\alpha }=(d\xi ,d\vartheta ,\delta \varphi ,\delta t),$ $%
\mathring{\psi}$ is the solution of
\begin{equation*}
\frac{\partial ^{2}\mathring{\psi}}{\partial \xi ^{2}}+\frac{\partial ^{2}%
\mathring{\psi}}{\partial \vartheta ^{2}}=\ ^{h}\mathring{\Lambda}.
\end{equation*}%
If the nonsymmetric part of (\ref{asol3}) is with $\mathbf{b}_{\alpha \beta
} $ being a solution of (\ref{ansymtors}), the rest of coefficients are
constrained to satisfy above mentioned conditions, we generate a class of
both nonholonomic and nonsymmetric metric deformations of the Schwarzschild
metric which defines a family of two parametric nonholonomic solutions in
NGT (when the gravitational field equations are approximated by (\ref{feq1})
and (\ref{feq1})). The stability conditions (\ref{stabcond}) result in
\begin{equation*}
\mathring{z}=g^{ij}\mathring{z}_{ij}(g_{ij,}g_{ab},,n_{i}^{a})=\ ^{h}%
\mathring{\Lambda}=\gamma \ \widehat{\mathbf{R}}_{\mu \alpha \nu \beta }%
\mathbf{b}^{\mu \nu }\mathbf{b}^{\alpha \beta }.
\end{equation*}%
This imposes a constraint of the generating function $q_{4}^{\hat{1}}(\xi
,\vartheta ,\varphi )$ and integration functions and constants, of type $%
q_{0}(r)$ and $\ ^{1}n_{i}(\xi ,\vartheta )$ and $\ ^{2}n_{i}(\xi ,\vartheta
),$ which selects of subspace in the integral variety of solutions of (\ref%
{aux3a}). We have $\mathring{z}=0$ and $\widehat{\mathbf{R}}_{\mu \alpha \nu
\beta },$ for any nonzero $\gamma,$ in the case of teleparallel nonholonomic
manifolds, see Chapter 1 in Ref. \cite{vsgg} (we note that for such
configurations the Riemann curvature for the Levi--Civita connection, in
general, is not zero).

We conclude that the presence of a small cosmological constant $\
^{h}\Lambda \approx \ \mathring{\varepsilon}^{2}\ ^{h}\mathring{\Lambda}$
may stabilize additionally the solutions but stability can be obtained also
for vanishing cosmological constants. Constructing such solutions we
considered, for simplicity, that the mass of effective gauge fields is very
small. In a more general case, we can generate nonsymmetric metrics with
effective Proca fields with nonzero mass and nonzero cosmological constants,
see more sophisticate constructions in Refs. \cite{vncg,vsgg,vrfg}.

\section{Conclusions and Discussion}

In this article we developed a new method of stabilization in nonsymmetric
gravity theories (NGT) and spacetimes provided with nonholonomic
distributions and canonically induced anholonomic frames with associated
nonlinear connection (N--connection) structures. For general effective
Lagrangians modelling NGT on (non) holonomic backgrounds, we shown how to
construct stable and nonstable solutions. We argued that the corresponding
systems of field equations possess different types of gauge like and
nonholonomically deformed symmetries which may stabilize, or inversely,
evolve into instabilities which depends on the type of imposed constrains
and ansatz for the symmetric and nonsymmetric components of metric and
related N--connection and linear connection structures.

The N--connection geometry and the formalism of parametric nonholonomic
frame transforms are the key prerequisites of the so--called anholonomic
frame method of constructing exact and approximate solutions in Einstein
gravity and various generalizations to (non)symmetric metrics,
metric--affine, noncommutative, string like and Lagrange--Finsler gravity
models, see reviews and explicit examples in Refs. \cite%
{ijgmmp,vrfg,vsgg,vncg}. Such geometric methods allow us to generate very
general classes of solutions of nonlinear field and constraints equations,
depending on three and four variables and on infinite number of parameters,
and solve certain stability problems in various models of gravity. For
simplicity, in this paper we consider the nonholonomic stabilization method
for a class of solutions with ellipsoidal symmetries which transform into
the Schwarzschild background for small eccentricities and small nonsymmetry
(of metrics) parameters.

The idea to use Lagrange multipliers and dynamical constraints proposed and
elaborated in Ref. \cite{moff5}, in order to solve instabilities discovered
in NGT by Clayton \cite{cl1,cl2}, contains already a strong connection to
the nonholonomic geometry and field dynamics. This work develops that
dynamical constraint direction to the case of nonholonomic parametric
deformations following certain results from Refs. \cite{avnsm01,avnsm02} (on
the geometry of generalized spaces and Ricci flows constrained to result in
nonholonomic and (non)symmetric structures). This way we can solve the
Janssen--Prokopec stability problem in NGT \cite{jp1,jp2,jp3} and develop a
new (nonholonomic) direction in (non) symmetric gravity and related
spacetime geometry. Here we also note that nonsymmetric components of
metrics arise naturally as generalized almost symplectic structures in
deformation quantization of gravity \cite{vqg4,avqg5} when corresponding
almost K\"{a}hler models are elaborated for quantum models. It was proved
how general relativity (GR) can be represented equivalently in nonsymmetric
almost symplectic variables for a canonical model on a corresponding almost K%
\"{a}hler spaces. For such a model of ''nonsymmetric'' gravity/ general
relativity, the questions on stability of solutions is to be analyzed as in
GR, together with additional considerations for nonholonomic constraints.

Following the above--mentioned results, we have to conclude that
nonsymmetric metrics and connections are defined naturally from very general
constructions in modern geometry, nonlinear functional analysis and
theoretical methods in gravity and particle physics. Such nonsymmetric
generalizations of classical and quantum gravity models can not be
prohibited by some examples when a gauge symmetry or stability scenaria fail
to be obtained for a fixed flat or curved background like in Refs. \cite%
{ddmcc,cl1,cl2,jp1,jp2,jp3}. It is almost sure that certain nonlinear
mathematical techniques always can be provided in order to construct stable,
or un--stable, solutions, with evolutions of necessary type; as well one can
be elaborated well defined physical scenaria and alternatives. This is
typical for generic nonlinear theories like GR and NGT.

Of course, there exists the so--called generality problem in NGT when a
guiding principle has to be formulated in order to select from nine and more
constants and extra terms in generalized Lagrangians (at least by 11
undetermined parameters come from the full theory and the decomposition of
the metric tensor). It may be that (non)symmetric corrections to metrics and
connections can be derived following certain geometric principles in Ricci
flow and/or deformation quantization theories, not only from the variational
principle for generalized field interactions and imposed nonholonomic
constraints. One also has to be exploited intensively certain variants of
selection from different theories following existing and further
experimental data like in \cite{moff2,moff3,mofft,jp3}, see also references
therein. At this moment, there are none theoretical and experimental
prohibitions for nonsymmetric metrics which would be established in modern
cosmology, astrophysics and experimental particle physics.

\vskip3pt

\textbf{Acknowledgement: } The work is performed during a visit at Fields
Institute.

\appendix

\setcounter{equation}{0} \renewcommand{\theequation}
{A.\arabic{equation}} \setcounter{subsection}{0}
\renewcommand{\thesubsection}
{A.\arabic{subsection}}

\section{Torsion and Curvature of d--Connections}

\label{append1}The torsion $\mathcal{T}$ of a d--connection $\mathbf{D}$ is
defined
\begin{equation*}
\mathcal{T}\mathbf{(X,Y)\doteqdot \mathbf{D}_{\mathbf{X}}Y-D}_{\mathbf{Y}}%
\mathbf{X-[X,Y],}
\end{equation*}%
for any d--vectors $\mathbf{X=}h\mathbf{X}+v\mathbf{X=\ }^{h}\mathbf{X}+\
^{v}\mathbf{X}$ and $\mathbf{Y=}h\mathbf{Y}+v\mathbf{Y,}$ with a
corresponding N--adapted decomposition into
\begin{eqnarray}
&&\mathcal{T}\mathbf{(X,Y)}=\{h\mathcal{T}(h\mathbf{X},h\mathbf{Y}),h%
\mathcal{T}(h\mathbf{X},v\mathbf{Y}),h\mathcal{T}(v\mathbf{X},h\mathbf{Y}),h%
\mathcal{T}(v\mathbf{X},v\mathbf{Y}),  \notag \\
&&\qquad {\ }\qquad v\mathcal{T}(h\mathbf{X},h\mathbf{Y}),v\mathcal{T}(h%
\mathbf{X},v\mathbf{Y}),v\mathcal{T}(v\mathbf{X},h\mathbf{Y}),v\mathcal{T}v%
\mathbf{X},v\mathbf{Y})\}.  \label{tors}
\end{eqnarray}%
The nontrivial N--adapted coefficients,
\begin{equation}
\mathcal{T}\mathbf{=\{T}_{~\beta \gamma }^{\alpha }=-\mathbf{T}_{~\gamma
\beta }^{\alpha }=\left(
T_{~jk}^{i},T_{~ja}^{i},T_{~jk}^{a},T_{~ja}^{b},T_{~ca}^{b}\right) \mathbf{%
\},}  \label{dtors}
\end{equation}%
can be computed by introducing $\mathbf{X=e}_{\alpha }$ and $\mathbf{Y=e}%
_{\beta }$ into (\ref{tors}), see details in Refs. \cite{avnsm02,vrfg}.

The curvature of a d--connection $\mathbf{D}$ is defined
\begin{equation}
\mathcal{R}\mathbf{(X,Y)\doteqdot \mathbf{D}_{\mathbf{X}}\mathbf{D}_{\mathbf{%
Y}}-D}_{\mathbf{Y}}\mathbf{D}_{\mathbf{X}}\mathbf{-D}_{\mathbf{[X,Y]}},
\label{curv}
\end{equation}%
The formulas for local N--adapted components and their symmetries, of the
d--torsion and d--curvature, can be computed by introducing $\mathbf{X}=%
\mathbf{e}_{\alpha },$ $\mathbf{Y}=\mathbf{e}_{\beta }$ and $\mathbf{Z}=%
\mathbf{e}_{\gamma }$ in (\ref{curv}). The nontrivial N--adapted
coefficients
\begin{equation}
\mathcal{R}\mathbf{=\{\mathbf{R}_{\ \beta \gamma \delta }^{\alpha }=}\left(
R_{\ hjk}^{i}\mathbf{,}R_{\ bjk}^{a}\mathbf{,}R_{\ hja}^{i}\mathbf{,}R_{\
bja}^{c}\mathbf{,}R_{\ hba}^{i}\mathbf{,}R_{\ bea}^{c}\right) \mathbf{\}}
\label{dcurv}
\end{equation}%
are given by formulas (\ref{dcurvc}), see details in Ref. \cite{avnsm02,vrfg}%
.

The simplest way to perform computations with d--connections is to use
N--adapted differential forms like $\ \mathbf{\Gamma }_{\ \beta }^{\alpha }=%
\mathbf{\Gamma }_{\ \beta \gamma }^{\alpha }\mathbf{e}^{\gamma }$ with the
coefficients defined with respect to (\ref{ddif}) and (\ref{dder}). For
instance, the N--adapted coefficients of torsion (\ref{tors}), i.e.
d--torsion, is computed in the form
\begin{equation*}
\mathcal{T}^{\alpha }\doteqdot \mathbf{De}^{\alpha }=d\mathbf{e}^{\alpha }+%
\mathbf{\Gamma }_{\ \beta }^{\alpha }\wedge \mathbf{e}^{\beta },
\end{equation*}%
where
\begin{eqnarray}
T_{\ jk}^{i} &=&L_{\ jk}^{i}-L_{\ kj}^{i},\ T_{\ ja}^{i}=C_{\ ja}^{i},\ T_{\
ji}^{a}=\Omega _{\ ji}^{a},\   \notag \\
T_{\ bi}^{a} &=&\frac{\partial N_{i}^{a}}{\partial y^{b}}-L_{\ bi}^{a},\
T_{\ bc}^{a}=C_{\ bc}^{a}-C_{\ cb}^{a},  \label{dtorsc}
\end{eqnarray}%
where $\Omega _{\ ji}^{a}$ is the curvature of N--connection (\ref{ncurv}).

By a straightforward d--form calculus, we can find the N--adapted components
of the curvature (\ref{curv}) of a d--connection $\mathbf{D},$
\begin{equation*}
\mathcal{R}_{~\beta }^{\alpha }\doteqdot \mathbf{D\Gamma }_{\ \beta
}^{\alpha }=d\mathbf{\Gamma }_{\ \beta }^{\alpha }-\mathbf{\Gamma }_{\ \beta
}^{\gamma }\wedge \mathbf{\Gamma }_{\ \gamma }^{\alpha }=\mathbf{R}_{\ \beta
\gamma \delta }^{\alpha }\mathbf{e}^{\gamma }\wedge \mathbf{e}^{\delta },
\end{equation*}%
i.e. the d--curvature,
\begin{eqnarray}
R_{\ hjk}^{i} &=&\mathbf{e}_{k}\left( L_{\ hj}^{i}\right) -\mathbf{e}%
_{j}\left( L_{\ hk}^{i}\right) +L_{\ hj}^{m}L_{\ mk}^{i}-L_{\ hk}^{m}L_{\
mj}^{i}-C_{\ ha}^{i}\Omega _{\ kj}^{a},  \notag \\
R_{\ bjk}^{a} &=&\mathbf{e}_{k}\left( L_{\ bj}^{a}\right) -\mathbf{e}%
_{j}\left( L_{\ bk}^{a}\right) +L_{\ bj}^{c}L_{\ ck}^{a}-L_{\ bk}^{c}L_{\
cj}^{a}-C_{\ bc}^{a}\Omega _{\ kj}^{c},  \notag \\
R_{\ jka}^{i} &=&e_{a}L_{\ jk}^{i}-\mathbf{D}_{k}C_{\ ja}^{i}+C_{\
jb}^{i}T_{\ ka}^{b},  \label{dcurvc} \\
R_{\ bka}^{c} &=&e_{a}L_{\ bk}^{c}-\mathbf{D}_{k}C_{\ ba}^{c}+C_{\
bd}^{c}T_{\ ka}^{c},  \notag \\
R_{\ jbc}^{i} &=&e_{c}C_{\ jb}^{i}-e_{b}C_{\ jc}^{i}+C_{\ jb}^{h}C_{\
hc}^{i}-C_{\ jc}^{h}C_{\ hb}^{i},  \notag \\
R_{\ bcd}^{a} &=&e_{d}C_{\ bc}^{a}-e_{c}C_{\ bd}^{a}+C_{\ bc}^{e}C_{\
ed}^{a}-C_{\ bd}^{e}C_{\ ec}^{a}.  \notag
\end{eqnarray}

Contracting the first and forth indices $\mathbf{\mathbf{R}_{\ \beta \gamma
}=\mathbf{R}_{\ \beta \gamma \alpha }^{\alpha }}$, one gets the N--adapted
coefficients for the Ricci tensor%
\begin{equation*}
\mathcal{R}\mathit{ic}\mathbf{\doteqdot \{\mathbf{R}_{\beta \gamma }=}\left(
R_{ij},R_{ia},R_{ai},R_{ab}\right) \mathbf{\}.}  \label{driccis}
\end{equation*}%
see explicit formulas in Ref. \cite{vrfg}. It should be noted here that for
general d--connections the Ricci tensor is not symmetric, i.e. $\mathbf{%
\mathbf{R}_{\beta \gamma }\neq \mathbf{R}_{\gamma \beta }.}$

Finally, we note that there are two scalar curvatures, $\ ^{s}R$ and $\ ^{s}%
\check{R},$ of a d--connection defined by formulas%
\begin{equation*}
\ ^{s}R=\mathbf{g}^{\beta \gamma }\mathbf{\mathbf{R}_{\beta \gamma }}%
\mbox{\
and \ }\ ^{s}\check{R}=\mathbf{\check{g}}^{\beta \gamma }\mathbf{\mathbf{R}%
_{\beta \gamma }.}
\end{equation*}%
Both geometric objects can be considered in generalized gravity theories.

Similar formulas holds true, for instance, for the Levi--Civita linear
connection $\bigtriangledown =\{\ _{\shortmid }\Gamma _{\beta \gamma
}^{\alpha }\}$ is uniquely defined by the symmetric metric structure (\ref%
{ansatz}) by the conditions $~\ _{\shortmid }\mathcal{T}=0$ and $%
\bigtriangledown \mathbf{g}=0.$ It should be noted that this connection is
not adapted to the distribution (\ref{whitney}) because it does not preserve
under parallelism the h- and v--distribution. Any geometric construction for
the canonical d--connection $\widehat{\mathbf{D}}$ can be re--defined by the
Levi--Civita connection by using the formula
\begin{equation*}
\ _{\shortmid }\Gamma _{\ \alpha \beta }^{\gamma }=\widehat{\mathbf{\Gamma }}%
_{\ \alpha \beta }^{\gamma }+\ _{\shortmid }Z_{\ \alpha \beta }^{\gamma },
\end{equation*}%
where the both connections $\ _{\shortmid }\Gamma _{\ \alpha \beta }^{\gamma
}$ and $\widehat{\mathbf{\Gamma }}_{\ \alpha \beta }^{\gamma }$ and the
distorsion tensor $\ _{\shortmid }Z_{\ \alpha \beta }^{\gamma }$ with
N--adapted coefficientswhere%
\begin{eqnarray}
\ _{\shortmid }Z_{jk}^{i} &=&0,\ _{\shortmid
}Z_{jk}^{a}=-C_{jb}^{i}g_{ik}h^{ab}-\frac{1}{2}\Omega _{jk}^{a},~_{\shortmid
}Z_{bk}^{i}=\frac{1}{2}\Omega _{jk}^{c}h_{cb}g^{ji}-\Xi
_{jk}^{ih}~C_{hb}^{j},  \notag \\
_{\shortmid }Z_{bk}^{a} &=&~^{+}\Xi _{cd}^{ab}~^{\circ }L_{bk}^{c},\
_{\shortmid }Z_{kb}^{i}=\frac{1}{2}\Omega _{jk}^{a}h_{cb}g^{ji}+\Xi
_{jk}^{ih}~C_{hb}^{j},  \label{deft} \\
\ _{\shortmid }Z_{jb}^{a} &=&-~^{-}\Xi _{cb}^{ad}~~^{\circ }L_{dj}^{c},\
_{\shortmid }Z_{bc}^{a}=0,\ _{\shortmid }Z_{ab}^{i}=-\frac{g^{ij}}{2}\left[
~^{\circ }L_{aj}^{c}h_{cb}+~^{\circ }L_{bj}^{c}h_{ca}\right] ,  \notag \\
\Xi _{jk}^{ih} &=&\frac{1}{2}(\delta _{j}^{i}\delta
_{k}^{h}-g_{jk}g^{ih}),~^{\pm }\Xi _{cd}^{ab}=\frac{1}{2}(\delta
_{c}^{a}\delta _{d}^{b}+h_{cd}h^{ab}),  \notag
\end{eqnarray}%
for$~^{\circ }L_{aj}^{c}=L_{aj}^{c}-e_{a}(N_{j}^{c}),$ are defined by the
generic off--diagonal metric (\ref{ansatz}), or (equivalently) by d--metric (%
\ref{shvm}) and the coefficients of N--connection (\ref{coeffnc}) \cite{vrfg}%
. If we work with nonholonomic constraints on the dynamics/ geometry of
gravity fields, it is more convenient to use a N--adapted approach. For
other purposes, it is preferred to use only the Levi--Civita connection.

\section{Redefinition of Constants}

In order to get a convenient form of effective Lagrangian, the constants
from (\ref{genlag}) and (\ref{decompm}) are re--defined in the form:
\begin{eqnarray}
\alpha  &=&\rho +\Xi -1,\ \beta =\frac{1}{2}(\frac{1}{2}-\rho +2\sigma ),\
\label{redefc} \\
\gamma  &=&\Xi =3\Sigma ^{2}\theta ^{2}\left[ 2(c_{1}+c_{3})+1-b_{2}\right]
+\Sigma ^{2}\left[ d-\frac{b_{2}}{3}+\frac{2}{3}(c_{1}+c_{3})-\frac{3}{8}%
\mathbb{L}^{2}\right]   \notag \\
&&+\Sigma \{2\left[ \theta (1+b_{1})-\frac{2}{3}-\frac{8a_{1}}{3}+\frac{a_{2}%
}{2}-\frac{b_{1}}{3}-\left( a_{1}+\frac{1}{2}\right) \frac{\mathbb{L}^{2}}{2}%
\right] -  \notag \\
&&\theta (\phi -\xi )\left[ 4(c_{1}+c_{3})-2b_{2}+2\right] \}  \notag
\end{eqnarray}%
for
\begin{eqnarray*}
\Sigma  &=&\frac{3a_{2}/2-b_{1}-8a_{1}-2}{b_{2}-2(c_{1}+c_{3})-3d_{1}},\
\mathbb{L=}\frac{2}{3}\frac{(1+2a_{1})[b_{2}-3d_{1}-2(c_{1}+c_{3})]}{%
2+8a_{1}-\frac{3}{2}a_{2}-b_{1}} \\
\mathbb{K} &\mathbb{=}&\frac{3(a_{1}-a_{2}/4)}{(1+2a_{1})}\mathbb{L+}d_{1}-%
\frac{b_{2}}{3}+\frac{2}{3}(c_{1}+c_{3}),\ \theta \equiv \frac{2\mathbb{K+L}%
}{\mathcal{A}-\mathcal{B}}, \\
\xi  &\equiv &\frac{(\mathcal{A}+3\mathcal{B})(b_{1}+1)}{\mathcal{A}^{2}+%
\mathcal{AB-B}^{2}},\ \phi =\psi =\frac{(\mathcal{A}+\mathcal{B})(b_{1}+1)}{%
\mathcal{A}^{2}+\mathcal{AB-B}^{2}}, \\
\mathcal{A} &=&2(1-b_{2}+c_{1}+c_{2}),\ \mathcal{B}=-2(c_{1}+c_{3}), \\
\ \Psi  &=&(b_{2}-1)(\xi -\phi )^{2}+2\phi (1+b_{1}),\ \Phi =\frac{1}{3}%
(\phi ^{2}+2\xi \phi )(b_{2}+c_{1}+c_{3}-1)^{2}, \\
\ \Omega  &=&(c_{1}+c_{3})(\xi -\phi )^{2}+\xi (1+b_{1}),
\end{eqnarray*}%
where the conditions
\begin{equation}
\Omega +3\Phi =-\frac{1}{4}\mbox{\ and \ }\Xi =\Psi -2\Omega   \label{aux4}
\end{equation}%
have to be imposed in order to get a stable effective Lagrangian in the flat
space limit.

\end{document}